\documentclass[fleqn,usenatbib]{mnras}

\usepackage[T1]{fontenc}
\usepackage{ae,aecompl}
\usepackage{multirow}

\DeclareRobustCommand{\VAN}[3]{#2}
\let\VANthebibliography\thebibliography
\def\thebibliography{\DeclareRobustCommand{\VAN}[3]{##3}\VANthebibliography}

\usepackage{graphicx}	
\usepackage{amsmath}	
\usepackage{amssymb}	
\usepackage{newtxtext,newtxmath}

\newcommand{\lya}{Ly$\alpha$ }
\newcommand{\mt}{m_{\rm turn}}

\newcommand{\appropto}{\mathrel{\vcenter{
  \offinterlineskip\halign{\hfil$##$\cr
    \propto\cr\noalign{\kern2pt}\sim\cr\noalign{\kern-2pt}}}}}

\title[Extracting reionization from Ly$\alpha$ forest]{Extracting the astrophysics of reionization from the Ly$\alpha$ forest power spectrum: a first forecast}

\author[P.~Montero-Camacho and Y.~Mao]{
Paulo Montero-Camacho\thanks{pmontero@tsinghua.edu.cn (PMC)}
and Yi Mao\thanks{ymao@tsinghua.edu.cn (YM)}
\\
Department of Astronomy, Tsinghua University, Beijing 100084, China\\
}

\date{Accepted 2021 September 6. Received 2021 August 27; in original form 2021 July 1.}

\pubyear{2021}

\begin{document}
\label{firstpage}
\pagerange{\pageref{firstpage}--\pageref{lastpage}}
\maketitle

\begin{abstract}
The impact of cosmic reionization on the Ly$\alpha$ forest power spectrum has recently been shown to be significant even at low redshifts ($z \sim 2$). This memory of reionization survives cosmological time scales because high-entropy mean-density gas is heated to $\sim 3\times10^4$ K by reionization, which is inhomogeneous, and subsequent shocks from denser regions. In the near future, the first measurements of the Ly$\alpha$ forest 3D power spectrum will be very likely achieved by upcoming observational efforts such as the Dark Energy Spectroscopic
Instrument (DESI). In addition to abundant cosmological information, these observations have the potential to extract the astrophysics of reionization from the Ly$\alpha$ forest. We forecast, for the first time, the accuracy with which the measurements of Ly$\alpha$ forest 3D power spectrum can place constraints on the reionization parameters with DESI. Specifically, we demonstrate that the constraints on the ionization efficiency, $\zeta$, and the threshold mass for haloes that host ionizing sources, $m_{\rm turn}$, will have the $1\sigma$ error at the level of $\zeta = 25.0 \pm 11.6$ and $\log_{10} (m_{\rm turn}/{\rm M}_\odot) = 8.7^{+0.36}_{-0.70}$, respectively.
The Ly$\alpha$ forest 3D power spectrum will thus provide an independent probe of reionization, probably even earlier in detection with DESI, with a sensitivity only slightly worse than the upcoming 21~cm power spectrum measurement with the Hydrogen Epoch of Reionization Array (HERA), 
i.e.\ $\sigma_{\rm DESI} / \sigma_{\rm HERA} \approx 1.5$ for $\zeta$ and $\sigma_{\rm DESI}/\sigma_{\rm HERA} \approx 2.0$ for $\log_{10}(\mt / $M$_\odot)$. Nevertheless, the Ly$\alpha$ forest constraint will be improved about three times tighter than the current constraint from reionization observations with high-z galaxy priors, i.e.\ $\sigma_{\rm DESI} / \sigma_{\rm current} \approx 1/3$ for $\zeta$.  

\end{abstract}




\section{Introduction}
\label{sec:intro}
In recent years, the \lya forest has been consolidated into a leading probe of the post-reionization epoch. Its sensitivity to neutral hydrogen gas clouds that track the underlying dark matter content has allowed for new cosmological parameter constraints \citep[e.g. eBOSS;][]{2019JCAP...07..017C}. Furthermore, soon the \lya forest 3D power spectrum will be measured with high signal-to-noise by the Dark Energy Spectroscopic Instrument \citep[DESI;][]{2016arXiv161100036D} and after that by the 4-metre Multi-Object Spectroscopic Telescope \citep[4MOST;][]{2019Msngr.175...50R}. Nevertheless, these new measurements also come with new challenging systematics. 
One of such systematics is the thermal relics from inhomogeneous reionization in the \lya forest, which impacts the evolution of small-scale structures \citep{2019MNRAS.487.1047M,2020MNRAS.499.1640M}. 


Cosmic reionization is one of the key milestones of the evolution of the Universe. During the dark ages, the Universe primarily consists of free streaming cosmic microwave background (CMB) photons and neutral hydrogen atoms. As structure formation proceeds, the first luminous objects form and start to ionize their surroundings. Ultimately, the resulting ionized bubbles will percolate the whole Universe at redshift $z \approx 6$ \citep[e.g.,][]{2015MNRAS.447..499M}. The midpoint of this process is currently constrained to very likely happen at $z \approx 7.7$ by the CMB optical depth measurement \citep{2020A&A...641A...6P}. Nevertheless, a delayed reionization  scenario is also allowed by increasing evidence 
\citep{2019MNRAS.485L..24K,2020MNRAS.491.1736K,2020MNRAS.494.3080N}, possibly due to the baryonic streaming velocities \citep{2021ApJ...908...96P}. 

The violent heating of the intergalactic medium (IGM), which reaches a temperature of $\gtrsim 10^4$ K during reionization, perturbs the IGM's temperature-density relation  \citep[e.g. ][]{2016ARA&A..54..313M}. However, as a consequence of the cooling rates, the conventional wisdom was that the IGM would have relaxed to the usual temperature-density power law by $z <4$ \citep[e.g.,][]{2019MNRAS.486.4075O,2019MNRAS.490.3177W}. Hence an effective mitigation strategy for the memory of reionization  -- especially for the 1D \lya power spectrum -- would be to focus on redshift bins $ z < 4$. However, \cite{2018MNRAS.474.2173H}, using high-resolution simulations, found a bi-modal temperature-density distribution, which consisted of a Low-Entropy High-Density mode, where the typical temperature-density relation can be recovered rather swiftly (i.e. consistent with a negligible memory of reionization at $z<4$), and a High-Entropy Mean-Density (HEMD) mode originating primarily from minivoids. In contrast to the conventional wisdom, the HEMD mode can lag behind the other mode and it can take longer to be relaxed into the usual temperature-density power law. Consequently, the imprint of patchy reionization survives to $z \sim 2$ (see Figure 5 of \citealt{2018MNRAS.474.2173H}).

Using the same high-resolution simulations, \cite{2019MNRAS.487.1047M} reported for the first time the impact of inhomogeneous reionization in the \lya forest 3D and 1D power spectrum at lower redshifts\footnote{The 1D results in \cite{2019MNRAS.487.1047M} were compromised by an error in their mapping from 3D to 1D that was later corrected in \cite{2020MNRAS.499.1640M}.}. The expected effect of the memory of inhomogeneous reionization in terms of the relative difference is currently estimated to range from six per cent at $z = 2$ up to fifty per cent at $z = 4$ for the 3D \lya power spectrum, and it ranges from a per cent at $z = 2$ up to roughly twenty per cent at $z = 4$ for the 1D power spectrum (\citealt{2020MNRAS.499.1640M}; hereafter ``MM20''). Given the strength of this effect, an enhanced mitigation strategy is needed. However, MM20 also showed the dependence of the memory of reionization on the astrophysics of reionization, and pointed out that, from the astrophysical point of view, this broadband systematic effect can be transformed into a window to the Epoch of Reionization (EoR), by exploiting different $k$-dependence of power spectrum among various astrophysical scenarios.

Here we further develop the method presented in MM20, and make a first forecast of the prospects of constraining the parameters of reionization model (hereafter ``reionization parameters'') with the post-reionization Ly$\alpha$ forest power spectrum. Specifically, we will utilize an ideal eBOSS-like instrument to make a forecast of reionization parameter accuracies using the imprint of thermal relics from patchy reionization on the \lya forest 1D power spectrum, and will employ the DESI spectrograph and quasar luminosity information to forecast the reionization parameter accuracies from the effect on the \lya forest 3D power spectrum. Note, however, that due to limitations in our approach, these forecasts are only intended as a tool to inform us how well DESI might extract the astrophysics of reionization, and they are by no means part of a precision cosmology program \emph{yet}. 

This paper is organized as follows. In \S\ref{sec:strat}, we describe the forecasting strategy. We report the forecast results using the \lya forest 1D power spectrum with an ideal eBOSS-like instrument in \S\ref{sec:1D}, and using the \lya forest 3D power spectrum with DESI in \S\ref{sec:3D}. 
We discuss the shortcomings and limitations of our approach in \S\ref{sec:disc}, and make concluding remarks in \S\ref{sec:conc}.


\section{Methodology}
\label{sec:strat}

\subsection{Simulations and forecast strategy}
The base \lya forest power spectra used throughout this work are modelled analytically, following \cite{2003ApJ...585...34M}. Our model includes the impact of the memory of reionization in the \lya forest that is based on numerical simulations first described in \cite{2019MNRAS.487.1047M}.
To account for long-lasting thermal relics from reionization, primarily sourced by HEMD gas that originates in minivoids and reionizes to high entropy, we choose to utilize a hybrid simulation approach wherein large-scale semi-numerical simulations take care of the inhomogeneous nature of reionization and small-scale high-resolution simulations carefully track the diffusion of neutral gas, in order to accommodate the large dynamical scales associated with the underlying physics. Specifically, our strategy for the  \lya forest consists of the ingredients as follows.

(1) \emph{Small-scale simulations:} we utilize the modified \texttt{Gadget2} \citep{2005MNRAS.364.1105S} simulations described and tested in detail in \cite{2018MNRAS.474.2173H}. These simulations with the size of $2551 \ \textup{ckpc}$, and with particle number of $2 \times (384)^2$, resolve the gas to below the Jeans mass prior to reionization. In these simulations, the dark matter particle mass is $9.72 \times 10^3 \ \textup{M}_\odot$ and the gas (particle) mass is $1.81 \times 10^3 \ \textup{M}_\odot$. Within this small volume, cosmic reionization is assumed to take place instantaneously at a given redshift $z_{\rm re}$. The output of these simulations is an optical depth cube, as a function of redshift of observation $z_{\rm obs}$, that compares how transmission in the \lya forest changes when reionization occurs (instantaneously) at $z_{\rm re}$ in comparison to when it happens at a pivot redshift $\bar{z}_{\rm re} \equiv 8.0$, i.e.\ $\tau_1 (z_{\rm obs},z_{\rm re}) / \tau_1 (z_{\rm obs},8.0)$. In order to increase the speed of these simulations, we have used $z_{\rm re} = 6.0$, $7.0$, $8.0$, $9.0$, and $12.0$, i.e. we dropped $10.0$ and $11.0$ compared to \cite{2019MNRAS.487.1047M}. We have checked that this speed-up only sacrifices a $\sim 5$ per cent in accuracy. After reionization, the particles are evolved with singly ionized physics for both \ion{H}{II} and \ion{He}{II}. There is no \ion{He}{II} reionization implemented. Likewise, there are no fluctuations in the low-redshift ionizing background implemented in these simulations. However, we do implement streaming velocities that modulate the amount of small-scale structure, and hence play a primary role in reducing the memory from reionization. These small-scale simulations are used to compute the transparency of the IGM, $\psi =\Delta \ln  \tau_1 (z_{\rm obs},z_{\rm re}) =  \ln [ \tau_1 (z_{\rm obs},z_{\rm re}) / \tau_1 (z_{\rm obs},\bar{z}_{\rm re} = 8.0)]$. 

(2) \emph{Large-scale simulations:} we use \texttt{21cmFASTv2} \citep{2011MNRAS.411..955M,2019MNRAS.484..933P} to account for the patchy nature of reionization. We choose a box size of $400 \ \textup{cMpc}$ to have enough statistical power in the wavenumber range of $0.03\le k \le 0.7 \, \textup{Mpc}^{-1}$. These simulations have $256^3$ cells and $768^3$ cells for the ionization and matter fields respectively. These semi-numerical simulations cover the redshift range of $5.90 \le z \le 34.7$. We choose to utilize a constant ionization efficiency and sharp cutoff for the mass of galaxies that release the UV photons responsible for reionization. These large-scale simulations are used to obtain the cross-correlation between matter and neutral fraction field for the duration of reionization, i.e.\ $P_{m,x_{\rm HI}}$ in Eq.~(\ref{eq:psi}) below, and thus we do not require snapshots from these simulations at the post-reionization epoch.   

Moreover, for the analytical modeling of the base \lya forest, the linear matter power spectrum of our cosmological models is constructed by employing the Boltzmann equation solver code \texttt{CLASS} \citep{2011JCAP...07..034B}.

This hybrid strategy is a functional pragmatic method, while it has some disadvantages as discussed in \S\ref{sec:disc}. For the purpose of our forecast, our small-scale high-resolution simulations are too slow to perform the Bayesian inference of cosmological parameters. Instead, we choose to employ the Fisher matrix formalism in this paper. Nevertheless, compared to the accuracy of MCMC sampling used in conjunction with the forecasting formalism, our forecast can point to the right order of magnitude and morphology of the parameter space, as long as degeneracies enter the likelihood linearly \citep{2012JCAP...09..009W}.  

With these simulations, and the analytical model from \cite{2003ApJ...585...34M}, we can calculate the \lya observable signals, $P_{\rm F}^{\rm 1D}$ and $P_{\rm F}^{\rm 3D,obs}$, and their derivatives. 
To construct the observed 3D \lya power spectrum, both a description of the spectrograph in the relevant band and the Quasar Luminosity Function (QLF) are required. The DESI collaboration kindly provided both of them. 
The QLF is based on the work made by \cite{2013A&A...551A..29P} and on the QSO target selection work of \cite{2020RNAAS...4..179Y}. There is no technical paper describing the spectrograph yet.

Another ingredient in Fisher matrix formalism is the uncertainty or variance (for the 3D case) of these observables. We use the statistical and systematic uncertainty (added in quadrature) reported by eBOSS in \cite{2019JCAP...07..017C} for the 1D \lya forest forecast (see \S\ref{ssec:eBOSS}). On the other hand, the methodology for the 3D \lya forecast, which is much more sophisticated, will be described in \S\ref{sec:3D}, following the procedure in \cite{2014JCAP...05..023F} and references therein.

\subsection{Parameter space}
\label{ssec:models}
Next, we describe the parameters -- and therefore models -- that are allowed to vary in this forecast.
We aim to forecast the astrophysics of reionization, which will be represented throughout this paper by the ionization efficiency ($\zeta$) and the mass threshold ($\mt$), marginalized over nuisance parameters ($\bar{F}$, $A_1$ and $n_1$) that are typically used to model the \lya forest. The full, 5-dimensional, parameter vector is given by $\boldsymbol{\theta} = \{ \bar{F}, A_1, n_1, \zeta, \mt \}$. 

(1) $\bar{F} = \langle \exp{(- \tau)} \rangle$, the observed mean transmitted \lya flux that governs over the \lya forest astrophysics. 

(2) $A_1= k_1^3 P_L(k_1) / (2\pi)^3$, the value of the dimensionless linear matter power spectrum evaluated at $k_1 \equiv 2 \pi\,h\,{\rm Mpc}^{-1}$. 

(3) $n_1$, the power-law index of the linear power spectrum evaluated at $k_1$. $A_1$ and $n_1$ contain cosmological information. 

(4) $\zeta$, the ionization efficiency that governs the amount of available UV photons that can ionize the \ion{H}{I} regions. Its primary role is to set the timing of reionization, e.g.\ the midpoint of reionization.

(5) $\mt$, the threshold (sharp cutoff) mass of haloes that host star-forming galaxies. This parameter expresses how difficult it is to find star-forming galaxies that contribute to UV photons at small haloes due to efficient atomic cooling. The main impact of $\mt$ is in the duration of reionization, although it can influence the midpoint of reionization too. 

We show the fiducial values and step sizes of our $\boldsymbol{\theta}$ vector in Table \ref{tab:params}. We have checked that our results do not depend strongly on the values of the step size of reionization parameters. Throughout this work we use the following fixed cosmology: $\Omega_b = 0.0486$, $\Omega_m = 0.3088$, $\Omega_{\Lambda} = 0.6912$, and $h = 0.6774$, while the values of $\sigma_8$ and $n_s$ are varied corresponding to $A_1$ and $n_1$ in Table~\ref{tab:params} (e.g. $\sigma_8 = 0.8159$, $n_s = 0.9667$ for the fiducial set).\footnote{Another often used parametrization of the cosmological information in the \lya forest is to use the dimensionless amplitude of the linear matter power spectrum $\Delta^2_{\rm L}(k_p,z_p)$ and the logarithmic slope evaluated at a pivot redshift $z_p$ and wavenumber $k_p$ at which the information is near maximum, $n_{\rm eff}(k_p,z_p)$  \citep{2005ApJ...635..761M,2019JCAP...07..017C}. For reference, our values of $\sigma_8$ and $n_s$ correspond to $\Delta^2_{\rm L}(k_p = 0.009\,{\rm s/km}, z_p = 3.0) = 0.36$ and $n_{\rm eff}(k_p,z_p) = -2.50$, respectively.}

For the nuisance parameters we follow the approach by \cite{2014JCAP...05..023F}, i.e. we base the modeling of these parameters on the values reported in Table 1 of \cite{2003ApJ...585...34M}. Note that the common main disadvantage of this approach is that their Table 1 was computed at low redshifts ($z = 2.25$), which implies that for higher redshifts the modeling will not be as reliable. This fact, coupled with the memory of reionization present at higher redshift, is part of the reason why the DESI forecasts limit the redshift range to $z =2$ -- $2.7$ \citep{2016arXiv161100036D}, which can be considered as a conservative choice. Thus, one of the main avenues for improvement in the future work is to utilize the results of \cite{2015JCAP...12..017A} that updates the \lya modeling, analogous to \cite{2003ApJ...585...34M}, up to $z=3$. Likewise, future \lya forecasts that focus on high redshift effects, particularly the memory of reionization, will benefit greatly from updating the work of \cite{2015JCAP...12..017A} to even higher redshifts.

Due to this crippling disadvantage, we will split our 1D analysis into two redshift ranges: the lower redshifts -- the more reliable range -- with $2.2 \le z_i \le 3.0$ and a high redshift cut to illustrate the most uncertain extrapolated regime, which will have redshift bins with $3.2 \le z_i \le 4.0$. Note that there is actually no gap between the low-z and high-z ranges since $z_i$ is just the central value of the $z$-bin.  This separation is also justified by the effects of \ion{He}{II} reionization in the thermal relics from \ion{H}{I} reionization, in particular on its effect on the HEMD phase of the temperature-density relation. Concerning the impact of \ion{He}{II} reionization, however, the reliable range is the high-redshift regime when it has not occurred yet or has not yet impacted the thermal evolution of the IGM significantly. Meanwhile, we do not consider the split in redshifts interesting in the 3D forecast, because we have chosen to treat the aliasing term (see \S\ref{sec:3D}) as pure noise that dominates the signal at high redshift. This forecasting choice makes our 3D forecast overall quite conservative. 


Readers familiar with typical \lya forecasts \citep[e.g.,][]{2014JCAP...05..023F} might wonder why we have chosen not to include the temperature-density relation $T = T_0 \Delta^{\gamma-1}$ as nuisance parameters in the forecast. The reason is that the memory of reionization primarily impacts the temperature-density relation; however, the work of \cite{2003ApJ...585...34M} does not account for this effect. Hence, we leave their evolution to the small-scale simulations, which include the HEMD phase of the temperature-density relation. This strategy is not ideal. However, we deem it good enough for this forecast due to the role of the HEMD gas in disrupting $\gamma$ and also because the derivative of the \lya forest power spectrum with respect to $T_0$ start to become non-negligible at larger wavenumbers than the ones we are interested (see Figure 10 (b) of \citealt{2003ApJ...585...34M}).

We do not vary the heating parameters of {\sc 21cmFAST} since they play a minor role on the impact of inhomogeneous reionization in the \lya forest (see MM20). Furthermore, we have assumed constant ionization efficiency and mass threshold. Although more realistic models can be constructed by letting both quantities evolve with halo mass (thus time) and including their evolution parameters in the forecast, our approach is sufficient for a first forecast of the memory of reionization in the \lya forest.

\begin{table}
\centering
\caption{Fiducial values and variations for the forecast parameters: the observed mean transmitted \lya flux ($\bar{F})$, the amplitude of the dimensionless linear matter power spectrum ($A_1$) and its tilt ($n_1$), the ionization efficiency ($\zeta$), and the threshold mass of haloes that host star-forming galaxies ($\mt$).}
\label{tab:params}
\begin{tabular}{ccc}
\hline\hline
Parameter $\boldsymbol{\theta}$ & Fiducial value & Step size $ \delta$\\
\hline
$\bar{F}$ & 0.844 & 0.05 \\
$A_1$ & 1.48 & 0.29 \\
$n_1$ & -3.19 & 0.10 \\
$\zeta$ & 25 & 1.0 \\
$\mt\,[\times 10^8\,{\rm M}_\odot]$ & $5.0$ & $1.0$ \\
\hline\hline
\end{tabular}
\end{table}

\begin{table*}
	\centering
	\caption{Transparency of the IGM -- estimated with the small-scale simulations -- for the different models used in the Fisher matrix computation. We tabulate how $\psi =\Delta \ln  \tau_1 (z_{\rm obs},z_{\rm re}) =  \ln [ \tau_1 (z_{\rm obs},z_{\rm re}) / \tau_1 (z_{\rm obs},\bar{z}_{\rm re} = 8.0)]$ changes for the different models, where $\tau_1$ is the optical depth needed for a patch of gas with temperature $10^4$ K and density $\Delta_b$ = 1 to reproduce the mean observed transmitted flux. Note that a negative value implies that the simulation that \emph{suddenly} reionized at $z_{\rm re}$ is more transparent than a reference simulation that suddenly reionized at $\bar{z}_{\rm re} =8.0$. Note that varying reionization parameters $\zeta$ and $\mt$ does not affect the value of $\psi$.}
	\label{tab:igm}
	\begin{tabular}{ccccccc}
		\hline\hline
		Model &  $z_{\rm obs}$ & $b_{\rm \Gamma}$ & \multicolumn{4}{c}{$\psi\times 10^5$ }  \\ \cline{4-7}
        {} & {} & {} & $z_{\rm re} = 6.0$ & $z_{\rm re} = 7.0$ & $z_{\rm re} = 9.0$ & $z_{\rm re} = 12.0$ \\
		\hline
		\multirow{5}{*}{\rotatebox[origin=c]{90}{Fiducial}} & 2.0 & 0.085 & 5702  & 1970 & -795.8  & -1440 \\ 
        {} & 2.5 & 0.146  & 5834  & 1900 & -401.4  & 923.5 \\
		{} & 3.0 & 0.235  & 8582  & 3057  & -661.4  & 1353 \\ 
        {} & 3.5 & 0.347  &  13705 & 5318  & -2108  & -1139 \\
        {} & 4.0 & 0.483  &  21721 & 8891  & -4265  & -5478 \\
        \hline
		\multirow{5}{*}{\rotatebox[origin=c]{90}{$\bar{F} - \delta \bar{F}$}} & 2.0 & 0.112 & 4838  & 1462 & -320.0  & -143.2 \\
        {} & 2.5 & 0.194  & 5855  & 1726 & -86.48  & 1936 \\
		{} & 3.0 & 0.307  & 10052  & 3407  & -714.2 & 1434 \\
        {} & 3.5 & 0.449  & 16869 & 6308  & -2539 & -2076 \\
        {} & 4.0 & 0.619  & 27050 & 10668 & -5152 & -7541 \\
        \hline
		\multirow{5}{*}{\rotatebox[origin=c]{90}{$\bar{F} + \delta \bar{F}$}} & 2.0 & 0.059 & 6828  & 2610 & -1382  & -2958 \\
        {} & 2.5 & 0.100  & 6191  & 2278 & -943.0  & -731.1 \\
		{} & 3.0 & 0.162  & 7314  & 2866  & -798.0  & 634.1 \\
        {} & 3.5 & 0.243  &  10300 & 4310  & -1758  & -572.8 \\
        {} & 4.0 & 0.342  &  15616 & 6930  & -3360  & -3516 \\
        \hline
		\multirow{5}{*}{\rotatebox[origin=c]{90}{$A_1 - \delta A_1$}} & 2.0 & 0.088 & 5421  & 1837 & -623.0  & -859.3 \\
        {} & 2.5 & 0.152  & 5844  & 1838 & -344.2  & 768.6 \\
		{} & 3.0 & 0.243  & 8841  & 2998  & -695.4  & 817.8 \\
        {} & 3.5 & 0.360  & 13948 & 5244  & -2124  & -1662 \\
        {} & 4.0 & 0.502  & 21981 & 8854  & -4250  & -5792 \\
        \hline
		\multirow{5}{*}{\rotatebox[origin=c]{90}{$A_1 + \delta A_1$}} & 2.0 & 0.084 & 6141  & 2304 & -1012  & -1826 \\
        {} & 2.5 & 0.143  & 5834  & 2133 & -433.2  & 913.6 \\
		{} & 3.0 & 0.228  & 8337  & 3100  & -669.8  & 1718 \\
        {} & 3.5 & 0.338  & 13519 & 5360  & -2161  & -762.6 \\
        {} & 4.0 & 0.468  & 21588 & 8952  & -4451  & -5346 \\
        \hline
        \multirow{5}{*}{\rotatebox[origin=c]{90}{$n_1 - \delta n_1$}} & 2.0 & 0.088 & 4979  & 1560 & -645.2  & -786.6 \\
        {} & 2.5 & 0.153 & 5659  & 1676 & -266.4  & 718.2 \\
		{} & 3.0 & 0.244  & 8905  & 2876  & -616.3  & 601.3 \\
        {} & 3.5 & 0.362  & 14367 & 5213  & -2060  & -1959 \\
        {} & 4.0 & 0.508  & 22894 & 8967  & -4221  & -6137 \\
        \hline
        \multirow{5}{*}{\rotatebox[origin=c]{90}{$n_1 + \delta n_1$}} & 2.0 & 0.083 & 6467  & 2584 & -1225  & -2410 \\
        {} & 2.5 & 0.142  & 5860  & 2211 & -683.2  & 788.4 \\
		{} & 3.0 & 0.226  & 8090  & 3219  & -911.3  & 1613 \\
        {} & 3.5 & 0.334 & 12847 & 5334  & -2405  & -837.0 \\
        {} & 4.0 & 0.462 & 20488 & 8708  & -4659  & -5459 \\
		\hline\hline
	\end{tabular}
\end{table*}

In Table \ref{tab:igm}, we tabulate the transparency of the IGM as a function of the redshift of observation and when the simulation instantaneously reionized for the different models utilized in this work, i.e.\ we tabulate $\psi =\Delta \ln  \tau_1 (z_{\rm obs},z_{\rm re}) =  \ln [ \tau_1 (z_{\rm obs},z_{\rm re}) / \tau_1 (z_{\rm obs},\bar{z}_{\rm re} = 8.0)]$, where $\tau_1$ is the optical depth that must be assigned to a patch of gas with mean density $\Delta_b = 1$ and temperature $T = 10^4$ K in order for the mean transmitted flux to match the observations. For our models that change astrophysics of reionization, we assume that they have the same IGM temperature evolution as the fiducial model. This is a consequence of our hybrid approach to the impact of inhomogeneous reionization in the \lya forest. Similarly, the models with different observed transmitted flux have the same inhomogeneous reionization results as the fiducial model, since $\bar{F}$ plays no role in computing the cross-power spectrum of matter and hydrogen ionization fraction. Naturally, if the dynamical scales of interest were not as large, one would like to have large-box high-resolution simulations capable of tracking HEMD gas, \lya forest physics, and inhomogeneous reionization. Thus, we deem our hybrid strategy acceptable for our forecasting purposes.

As should be expected, Table \ref{tab:igm} shows that when the observed flux increases, the radiation bias, $b_{\Gamma} = \partial \ln \bar{F} / \partial \ln \tau_1$, decreases since the map from $\tau_1$ to flux requires less radiation to match the observed flux at all redshifts -- and this is more dominant than the overall increase in transmitted flux. Moreover, an increase in the amplitude of the linear matter power spectrum, i.e. $A_1$ (and therefore $\sigma_8$), leads to more structures and thus less radiation bias needed. Similarly, increasing $n_1$, i.e. increasing $n_s$, also leads to a decrease in the radiation bias needed to recover the observed transmitted flux. 

Accounting for the relatively large expected scatter (e.g.\ see the error values included in Table 3 of \citealt{2019MNRAS.487.1047M}) of the IGM results obscures the meaning of our transparency results. Nevertheless, this scatter becomes less obscure for the derivative of transparency of the IGM with respect to redshift of sudden reionization, which is the quantity needed to compute the memory of reionization in the \lya forest. In general, the change of the transparency of the IGM with respect to $z_{\rm re}$ decreases faster at low $z_{\rm obs}$ for models with more structures because of more HEMD gas, i.e. the absolute value of $\partial \psi (z_{\rm obs}, z_{\rm re}) / \partial z_{\rm re}$ is larger for both an increase of $A_1$ or an increase of $n_1$ at low redshift of observation. 

Moreover, a scenario with less observed transmitted flux leads to an increase -- in absolute value -- of the transparency of the IGM at high redshifts. This behavior is due to modifying the overall normalization of the \lya opacity cube from our simulations, which is then used to match our transparency to the mean observed flux. In other words, a ``lesser forest'' -- larger $\bar{F}$ -- leads to a considerably more steep change of transparency of the IGM at high redshifts. This behavior is quickly subdued due to less discrepancy between the mean observed fluxes as redshift decreases and because the scenario with smaller $\bar{F}$ has more thermal relics to dissipate at lower redshifts since they survive longer.

\section{Forecast with 1D power spectrum}
\label{sec:1D}

\subsection{Modeling \lya forest 1D power spectrum}

In contrast to the 3D counterpart, the 1D \lya power spectrum $P_{\rm F}^{\rm 1D}$ has already been measured with high signal-to-noise \citep{2019JCAP...07..017C}, but there are still more upcoming efforts for higher resolution measurements \citep[e.g. WEAVE-QSO;][]{2016sf2a.conf..259P}. However, the effect of inhomogeneous reionization is comparable to the statistical error reported by eBOSS at the lower redshifts (\citealt{2019MNRAS.487.1047M}; MM20). Hence in order to attempt to extract the astrophysics of reionization from the 1D \lya forest, one would require an ideal eBOSS-like instrument with no systematic errors. For example, WEAVE's science goals include the search for thermal relics from hydrogen reionization. The need for an ideal eBOSS-like instrument, coupled to the fact that our approach ignores other systematics (e.g. clustering of UV fluctuation \citep{2014PhRvD..89h3010P}, AGN feedback \citep{2020MNRAS.495.1825C} and the imprint of \ion{He}{II} reionization \citep{2020MNRAS.496.4372U}), lead us to do a forecast for the 1D \lya forest that might be simplistic, instead of a full parameter inference, which we leave to future work. 
However, such a simple forecast is still valuable since DESI and WEAVE will measure the 1D \lya power spectrum. Also, understanding the trends in the 1D \lya power spectrum will allow us to build intuition before tackling the more complicated 3D \lya power spectrum.   

In this section, our observable is the \emph{total} 1D \lya forest power spectrum with the form \citep{2019MNRAS.487.1047M}
\begin{eqnarray}
    \label{eq:P1D}
    P_{F}^{\rm 1D} (z,k)  = b_{F}^2 (z) P^{\rm 1D}_{m} (z,k) + 2 b_{F} (z) b_{\rm \Gamma} (z) P_{m,\psi}^{\rm 1D} (z,k) \, \textup{,}
\end{eqnarray}
where the first term corresponds to the \emph{conventional} \lya forest power spectrum and it is computed using the fits from Table 1 of \cite{2003ApJ...585...34M}. The second term is the contribution due to the memory of reionization which is constructed using the expression \citep{2019MNRAS.487.1047M}
\begin{eqnarray}
    P_{m,\psi}^{\rm 1D} = \int^{\infty}_0 \frac{dk_\perp}{2 \pi} k_\perp (1 + \beta_F \mu^2) P_{m,\psi} (k,z_{\rm obs}) \, \textup{,} \\
    \label{eq:psi}
    P_{m,\psi}(k,z_{\rm obs}) = - \int_{z_{\rm min}}^{z_{\rm max}} dz \frac{\partial \psi}{\partial z}(z_{\rm obs},z) P_{m,x_{\rm HI}}(k,z)\frac{D_g(z_{\rm obs})}{D_g(z)}\, .
\end{eqnarray}
The integration in Eq.~(\ref{eq:psi}) covers the epoch of reionization, and we set the lower limit of integration $z_{\rm min} = 5.90$ and the upper
limit $z_{\rm max} = 34.7$. $D_g$ is the growth factor. $\partial \psi/\partial z$ quantifies how the transparency of the IGM changes when a patch of the sky reionizes at a different redshift, which is computed from our small-scale simulation results. $P_{m,x_{\rm HI}}$ is the cross-power between matter and neutral fraction field, which is computed from our large-scale simulation results.

For the effect of redshift-space distortion (RSD), we take a pragmatic approach, i.e. multiplying $P^{\rm 1D}_{m} (z,k)$ by $[(1 + z)/(1+z_{\rm ref})]^{3.55}$,  based on Eq.~(14) of \cite{2013A&A...559A..85P} throughout this paper, instead of choosing a constant RSD parameter $\beta_{F} = 1$ and varying $b_{F}(z)$ as in our previous work \citep{2019MNRAS.487.1047M}. Specifically, we choose $\beta_{F} = 1.58$, $b^2_{F} = 0.0173$, for our fiducial values of nuisance parameters, and select a reference redshift $z_{\rm ref}=2.25$ to compute the matter power spectrum. Moreover, $P_m^{\rm 1D}$ includes the non-linear correction $D(z,k,\mu)$ (not to be confused with the growth factor) described in Eq.~(21) of \cite{2003ApJ...585...34M}. Note that the values of both $\beta_{F}$ and $b_{F}$ change for the models that vary the nuisance parameters, as tabulated in Table 1 of \cite{2003ApJ...585...34M}.

\begin{figure*}
    \centering
    \includegraphics[width=\textwidth,keepaspectratio]{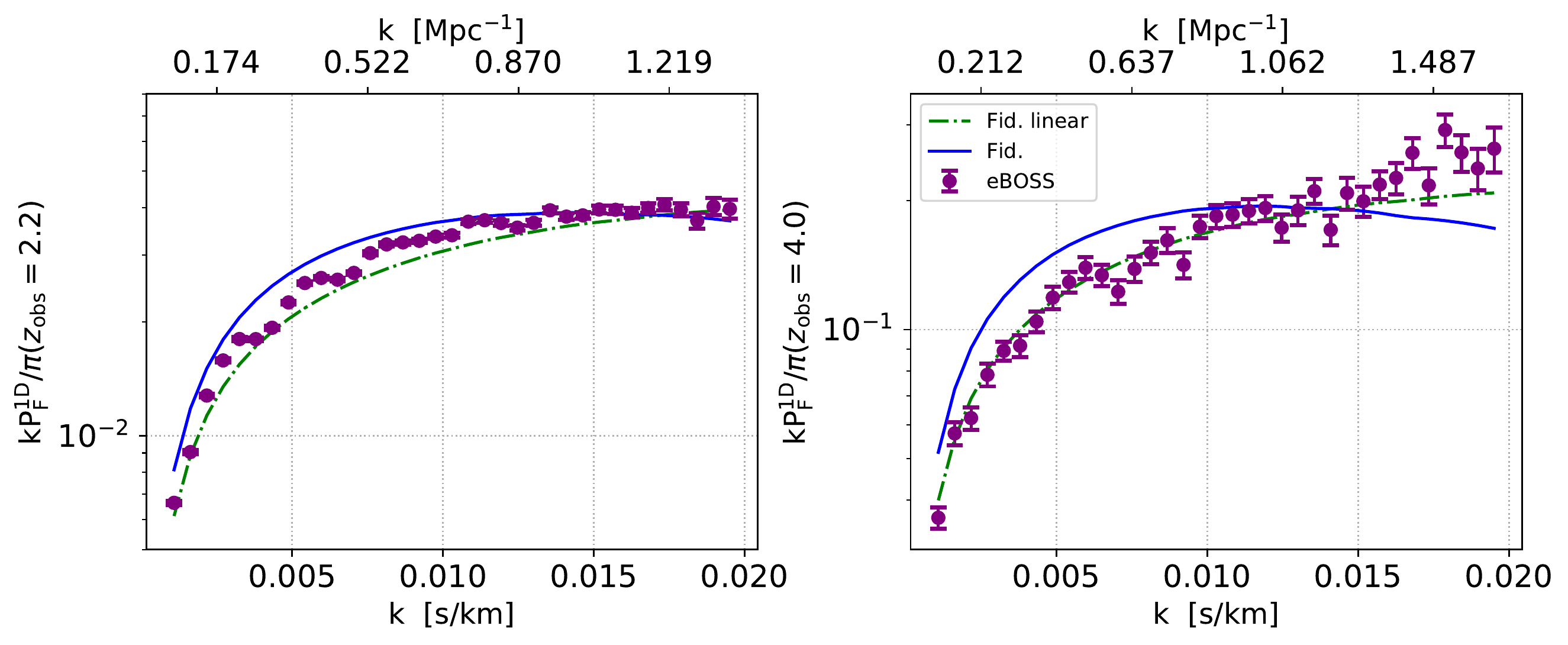} 
    \caption{Total (nonlinear \lya and reionization) 1D \lya forest power spectrum at the lowest redshift bin, $z=2.2$ (left), and at the highest redshift bin, $z = 4.0$ (right), for our fiducial model (blue solid line). For comparison, we show the recent eBOSS measurements \citep{2019JCAP...07..017C} (purple circles), and the total \emph{linear} \lya forest power spectrum (green dot-dashed line). For convenience, we mark the wavenumber units both in the \lya convention (lower axis) and in the cosmological convention (upper axis).  }
    \label{fig:P1D}
\end{figure*}

\begin{figure*}
    \centering
    \includegraphics[width=0.99\textwidth,keepaspectratio]{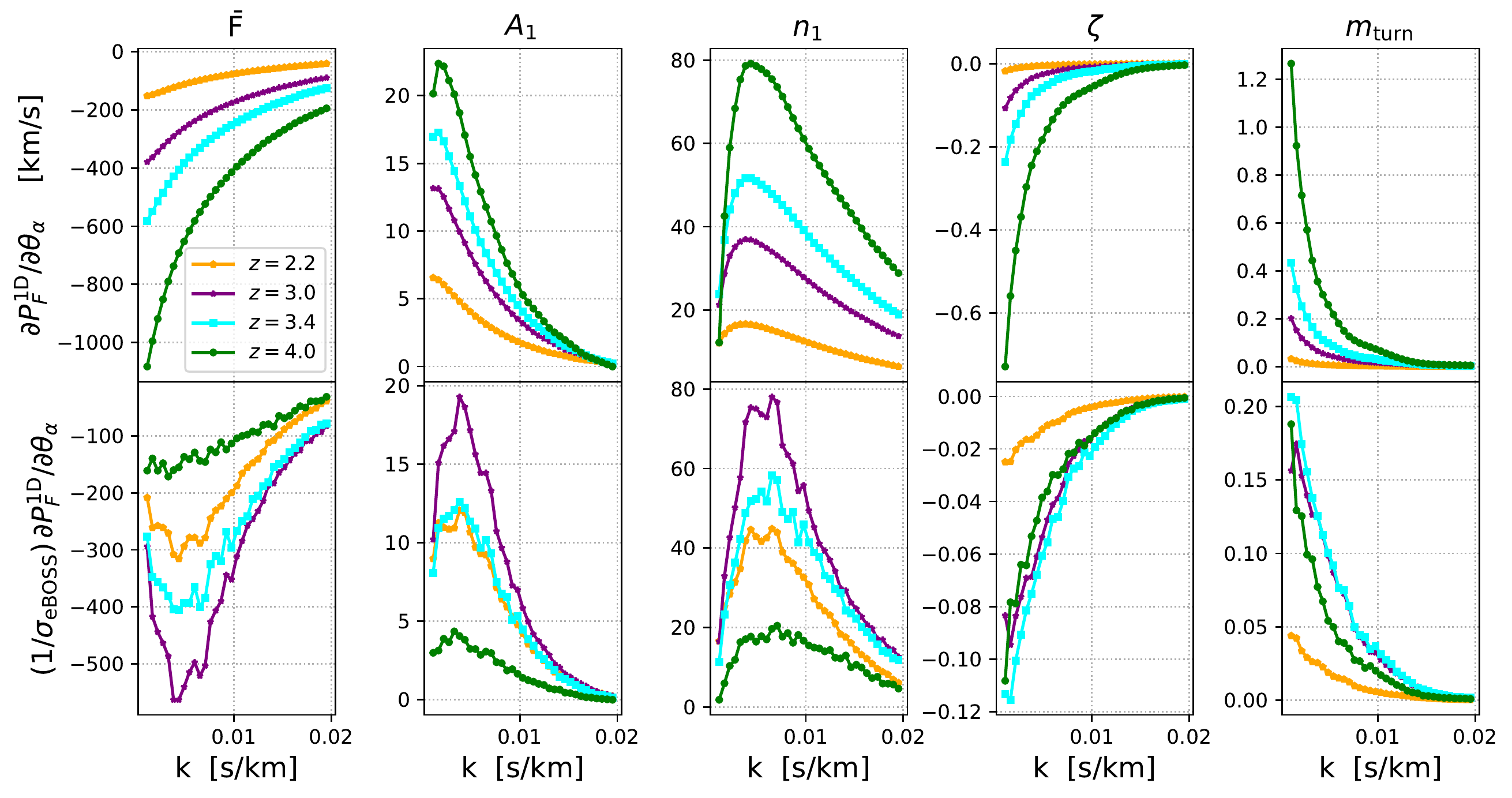}
    \caption{(Top) derivatives of the \emph{total} \lya forest 1D power spectrum $\partial P^{\rm 1D}_{F}/\partial \theta_\alpha$ (in units of ${\rm km/s}$) with respect to parameter $\theta_\alpha$, where the parameter is the (from the left to right) the observed transmitted flux $\bar{F}$, amplitude of linear power spectrum $A_1$, tilt of the linear power spectrum $n_1$, ionization efficiency $\zeta$, and threshold mass $\mt$. Shown are the results at different redshift $z=2.2/3.0/3.4/4.0$ (orange pentagons/purple stars/cyan squares/green circles), respectively. 
    (Bottom) derivatives weighted by our assumed error bars, i.e.\ $(1/\sigma_{\rm eBOSS})(\partial P^{\rm 1D}_{F}/\partial \theta_\alpha)$.}
    \label{fig:partial_P1D}
\end{figure*}

We plot the total 1D \lya forest power spectrum for our fiducial model in Figure \ref{fig:P1D}. Our fiducial model power spectrum agrees with the observational data \citep{2019JCAP...07..017C} reasonably well at both low and high redshifts. However, the inclusion of pixelization\footnote{Pixelization is added via multiplying $P_{F}^{\rm 1D}$ by a sinc kernel $\sin{x}/x$, where $x = 0.5 k \Delta v$ and $\Delta v = 50 $ km/s is the pixel width in velocity space.} (consistent with the expected performance of the DESI spectrograph) and nonlinearities, e.g. pressure correction and finger of god effect, have made our fiducial model slightly higher than the previous results of MM20. Hence we also plot the linear power spectrum with pixelization to illustrate this fact.


The Fisher matrix for our observable, $P_{F}^{\rm 1D}$, assuming that different bins are independent, is given by
\begin{eqnarray}
    \label{eq:fish1D}
    F_{\alpha \beta} = \displaystyle \sum_{i}^{z-{\rm bins}}\sum_{j}^{k-{\rm bins}} \sigma^{-2}_{z_ik_j}(P^{\rm 1D}_{F}) \frac{\partial P^{\rm 1D}_{F}}{\partial \theta_\alpha} (z_i,k_j)  \frac{\partial P^{\rm 1D}_{F}}{\partial \theta_\beta}(z_i,k_j) \, \textup{,}
\end{eqnarray}
where $\sigma_{z_ik_j}$ is the uncertainty of the corresponding bin, and $\theta_i$ is our parameter vector, i.e. $\theta_i = \{\bar{F}, A_1, n_1, \zeta, \mt\}$. We normalize $\mt$ by $10^8$ M$_\odot$ for numerical convenience. We obtain $\sigma$ by adding in quadrature the statistical and systematic  errors tabulated in \cite{2019JCAP...07..017C}. 

We show both the derivatives and weighted derivatives of our observable with respect to the forecast parameters in Figure~\ref{fig:partial_P1D}. All parameters show the expected increase (in absolute value) of the derivatives with larger redshift due to larger $P_{\rm F}^{\rm 1D}$. However, as noise gets included in the analysis, we see that the decrease of quasars lines of sight at larger redshifts leads to a decrease of the impact of the derivatives at large redshift and similarly a larger weight of the results for lower redshifts in the forecast. 

In addition, in Figure \ref{fig:partial_P1D} we also see the expected signs for the derivatives (and weighted-derivatives), i.e. if increasing a parameter lowers the power spectrum then the derivative is negative. This is the case for the mean transmitted flux since an increase leads to less absorption and therefore a ``less abundant'' forest. Likewise, a larger value of $\zeta$ leads to more UV photons in the IGM that can reionize neutral hydrogen regions, and hence it leads to early reionization models and consequently a smaller memory of reionization in the forest. In contrast, the parameters that lead to a positive derivative are $A_1$, $n_1$, and $\mt$. In the case of $A_1$, the reason for the sign is that an increase in $A_1$ translates into a larger value of $\sigma_8$ (alternatively of the amplitude of the curvature power spectrum), and thus more neutral hydrogen clouds will be able to absorb \lya radiation from distant quasars. Analogously, an increase in $n_1$ leads to an increase in the tilt of the curvature power spectrum, which translates into an increase in the number of neutral hydrogen clouds that form and survive to absorb \lya radiation. Likewise, an increase in the tilt will lead to more UV sources forming at small scales, and thus an earlier reionization at the cost of lower strength in the matter power spectrum at large scales. However, as mentioned in \S\ref{ssec:models}, an increase in $n_1$ can also lead to a less transparent IGM. Finally, for $\mt$ a larger value delays reionization since it is more difficult to form star-forming galaxies, and hence it leads to a stronger memory of reionization.

\begin{figure*}
    \centering
    \includegraphics[width=\textwidth,keepaspectratio]{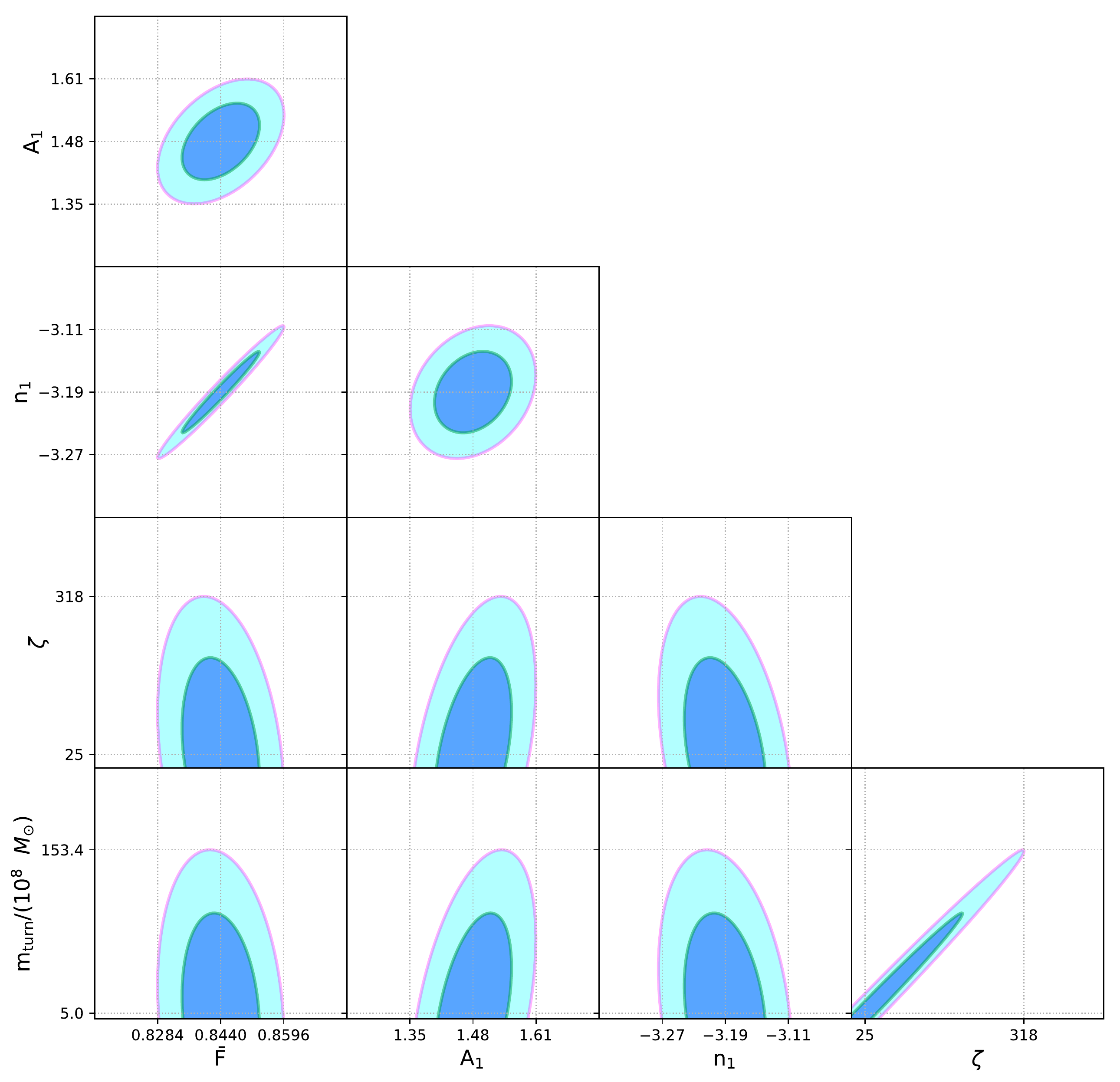}
    \caption{Forecast of parameter estimation from the 1D \lya power spectrum for an ideal eBOSS-like instrument with uncertainty corresponding to adding in quadrature the statistical and systematical errors of eBOSS, assuming a Low-z survey with redshift range $2.2 \leq z \leq 3.0$ and bin size $\Delta z = 0.2$. All contour plots have been centered on our fiducial values.  The blue and cyan ellipses correspond to the 1$\sigma$ and 2$\sigma$ contours, respectively.}
    \label{fig:eBOSS_lowz}
\end{figure*}

\begin{figure*}
    \centering
    \includegraphics[width=\textwidth,keepaspectratio]{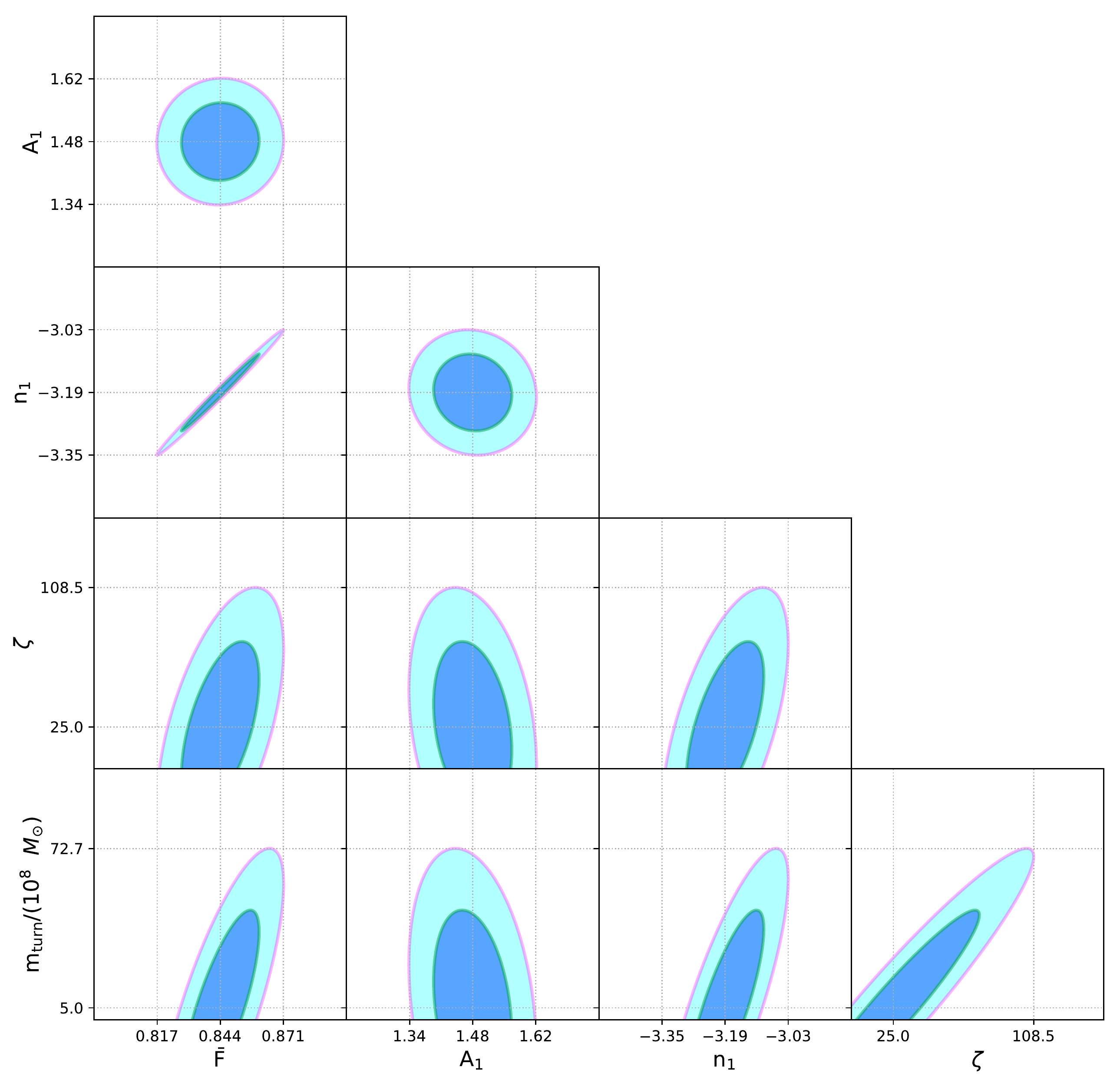}
    \caption{Same as Figure \ref{fig:eBOSS_lowz} but assuming a High-z \lya survey with redshift range $3.2 \leq z \leq 4.0$.} 
    \label{fig:eBOSS_highz}
\end{figure*}

Furthermore, in Figure \ref{fig:partial_P1D} we see the suppression of the \lya flux power spectrum at small scales due to the nonlinear corrections, which translates into a decrease of the derivative with respect to $A_1$. Besides, we see the expected increase in large-scale power due to an increase in the amplitude of the linear matter power spectrum and also due to the isotropic increase in power seeded by nonlinear growth. Similarly, for $n_1$ we also see a larger 1D flux power at larger scales; however, it falls quickly at even smaller $k$ due to the smaller amplitude of the total flux power spectrum at large scales for increasing $n_1$. Moreover, we see a relatively stronger impact of the nonlinear growth since its role is enhanced for $n_1$ (see Table 1 of \citealt{2003ApJ...585...34M}). We will see in \S\ref{sec:3D} that the explanations for the cosmological nuisance parameters are incomplete, since integrating along the perpendicular direction wipes-out the anisotropic features.  


Likewise, Figure \ref{fig:partial_P1D} has the expected hierarchy for the $\zeta$ and $\mt$ models, i.e. larger flux power at higher redshifts and large scales, where the memory of reionization is stronger because the IGM has not had enough time to relax into the usual temperature-density relation. 

Also, as expected, Figure \ref{fig:partial_P1D} shows that the nuisance parameters are the more sensible ones. Particularly, the observed transmitted flux naturally is the most important parameter since it directly determines the strength of the \lya forest spectra. Although the derivatives for the reionization parameters are low, a direct comparison would lead to underestimating their role in the Fisher matrix computation, i.e. the astrophysics of reionization play an ``accompanying'' role instead of the ``main melody''. However, we highlight that the increase in uncertainty at high redshift weights-down the most sensitive part of the forest to the astrophysics of reionization. 

We note that while several improvements could be made to our \lya forest modeling, our modeling is capable of reproducing the current state-of-the-art observations of the 1D \lya power spectrum, and therefore reasonably good enough for forecasting purpose. 


\subsection{Forecast for an ideal eBOSS-like survey}
\label{ssec:eBOSS}
In this subsection, we show the results obtained by utilizing an \emph{ideal} eBOSS-like instrument, with statistical and systematic errors equal to that of eBOSS. Given the observational uncertainty that is comparable to the effect of patchy reionization, the ``ideal'' survey means that one can effectively separate the astrophysics of reionization completely from all other systematic errors that are extracted before obtaining $P_{F}^{\rm 1D}$, which is quite an unrealistic scenario. Therefore, the objective here is to forecast how well the memory of inhomogeneous reionization might be ``seen'' by DESI and WEAVE-QSO in the near-future only with information of the 1D \lya forest power spectrum, and the main motivation is to build intuition for the more sophisticated scenario using the 3D \lya power spectrum. 

As mentioned in \S\ref{sec:strat}, we split our analysis into two different hypothetical surveys --- Low-z and High-z --- due to the absence of \ion{He}{II} reionization, which can cause a loss of sensitivity to the memory of hydrogen reionization, and because the way we model our nuisance parameters also leads to uncertainties regarding the range of validity of the nuisance-model at higher redshifts. The first survey, Low-z, covers the redshift range $2.2 \leq z \leq 3.0$ with the bin size $\Delta z = 0.2$. Hence this survey is robust with respect to the validity of the nuisance parameters, but not very reliable concerning the astrophysics of reionization since it only covers the range where \ion{He}{II} reionization is important. The second survey, High-z, covers $3.2 \leq z \leq 4.0$ with the same binning. This second survey is robust regarding the impact of thermal relics in the \lya forest since it stays away from where helium will complicate the thermal evolution of the IGM. However, regarding the modeling of the nuisance parameters, this survey can be simplistic. Both surveys adopt the same $k$-bins used in \cite{2019JCAP...07..017C}.

In Figure \ref{fig:eBOSS_lowz}, we show the corner plot for the forecast using the Low-z survey. 
The error-ellipses are centered at the fiducial values. As such, the shape and size of the contour are the interesting features in our forecasts. As highlighted by Figure \ref{fig:partial_P1D}, the errors on the nuisance parameters, particularly the error on the observed flux, are the most heavily constrained. In contrast, the reionization parameters are poorly constrained for this redshift range. Moreover, we find that the reionization parameters are only slightly degenerate with the nuisance parameters but heavily degenerate with themselves. This heavy degeneracy between $\mt$ and $\zeta$ is expected since increasing both parameters simultaneously influence the reionization history in opposite ways. The shape of their (weighted) derivatives are similar, which implies that a change in the observable due to one parameter can be countered by a move in the other parameter. 


Likewise, in Figure \ref{fig:eBOSS_highz} we plot the results of our forecast for the hypothetical High-z survey. Overall, we see a reduction of the contour area for reionization parameters. This can be attributed to two main factors. First, the memory of reionization in the 1D \lya forest power spectrum is a factor of a few larger in this redshift range compared to the one covered by Low-z. As a result, we are able to better forecast the reionization parameters even with potentially less available data. Secondly, this range is the most uncertain for modeling the nuisance parameters that uses $z = 2.25$ as the pivot redshift. Thus, the contours might be smaller than the true contours due to deviations in the values of the non-linear corrections, RSD parameter, and flux bias coefficient. Furthermore, we see the same expected degeneracies between $\bar{F}$ and $n_1$; however, the error on $A_1$ appears not to  degrade as much for these redshift bins compared to the other nuisance parameters. We attribute this behavior to the gains in the constraining power due to the role of the amplitude of the linear matter power spectrum on the memory of reionization. We also see some degeneracies between $\mt$ and the nuisance parameters. In particular, the degeneracy with $A_1$ and $n_1$ highlights the fact that all of these parameters play a role in governing the number of UV sources that will be present at a given redshift. Similarly, we see a strong degeneracy between $\zeta$ and $\mt$. Interestingly, the High-z constraints lead to a higher angle for the $\zeta$-$\mt$ ellipse ($\theta_{\rm High-z} = 38.7^{\circ}$ vs $\theta_{\rm Low-z} = 26.8^{\circ}$). This feature of the High-z forecast is due to the fact that High-z constrains the ionization efficiency better than the Low-z survey. We will see this is a unique feature of $P_F^{\rm 1D}$ and it implies that High-z is more sensitive to the midpoint of reionization than Low-z, since ionization efficiency primarily controls the timing of reionization (see MM20).

The significant improvement on the constraints of reionization parameters using only high-redshift bins is an encouraging sign for splitting future \lya Bayesian parameter inference into a cosmological inference at lower redshift bins and an EoR-focused one at the highest bins.

\begin{figure*}
    \centering
    \includegraphics[width=\textwidth,keepaspectratio]{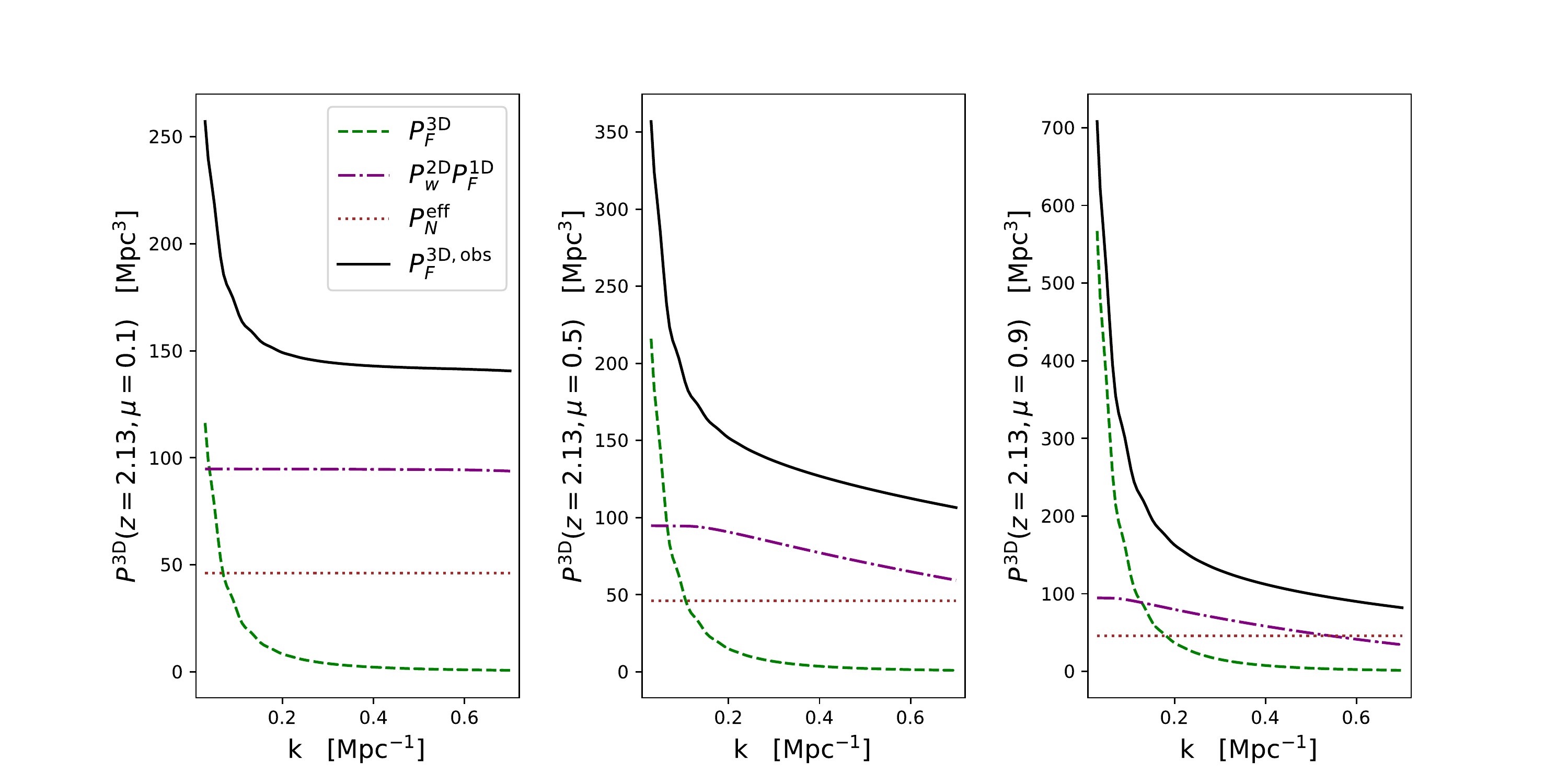}
    \caption{\emph{Total} observed \lya 3D power spectrum (black solid) and its components --- the \lya 3D power spectrum (green dashed), the aliasing term (purple dot-dashed), and the effective noise (red dotted), as a function of  wavenumber $k$. Shown are the power spectra at $z=2.13$ for the fiducial model, at a given $\mu = \cos\Theta= 0.1/0.5/0.9$ (from left to right) where $\Theta$ is the angle of $\boldsymbol{k}$ with respect to the line of sight. }
    \label{fig:P3D_total}
\end{figure*}

\begin{figure*}
    \centering
    \includegraphics[width=\textwidth,keepaspectratio]{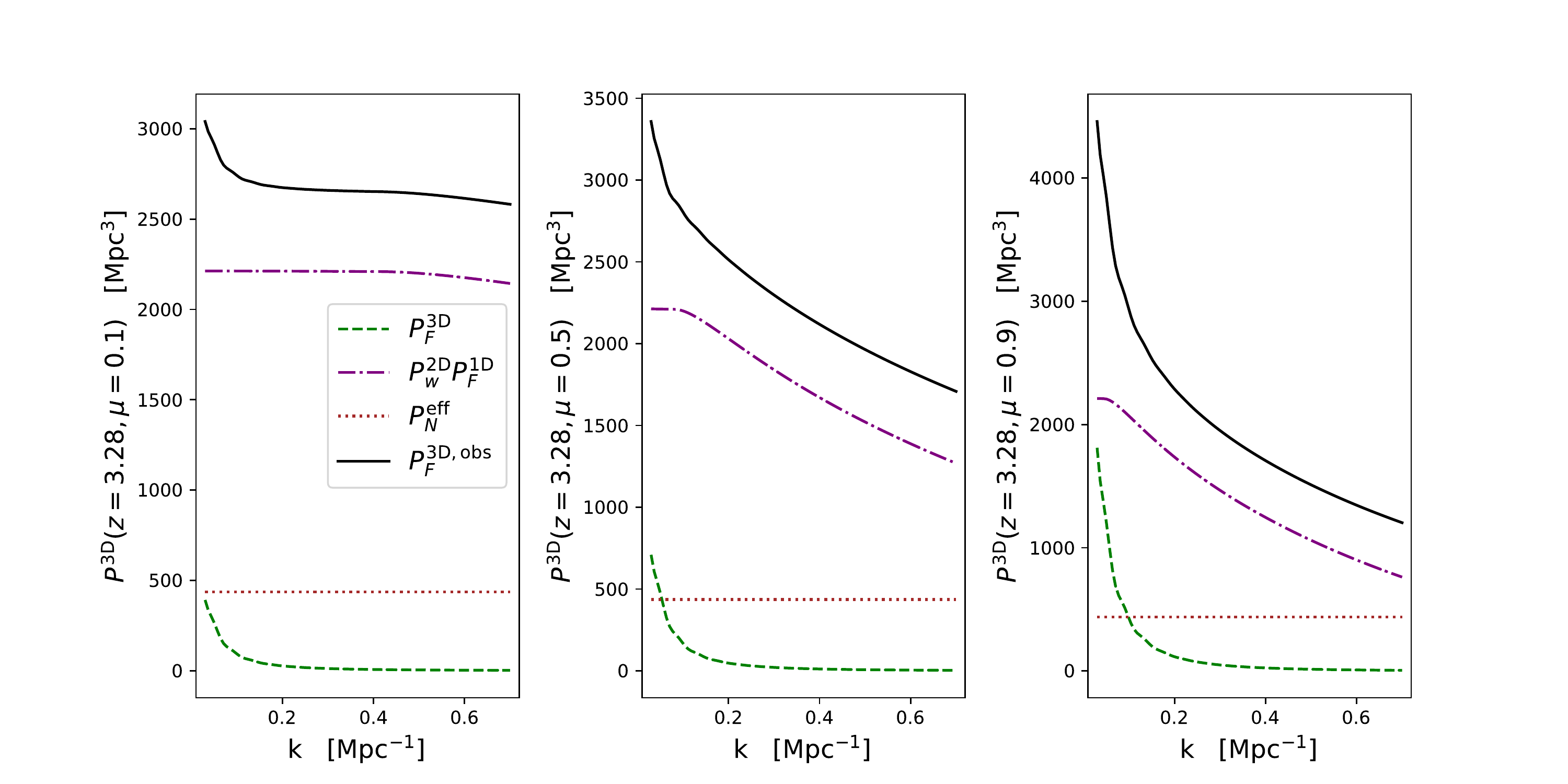}
    \caption{Same as Figure \ref{fig:P3D_total} but at $z = 3.28$.}
    \label{fig:P3D_total_II}
\end{figure*}

\section{Forecast with 3D power spectrum}
\label{sec:3D}

\subsection{Modeling \lya forest 3D power spectrum}

Now we turn to focus on the much more convoluted forecast with the 3D \lya forest power spectrum. 
We follow the approach used in \cite{2014JCAP...05..023F}. The Fisher matrix using 3D \lya forest power spectrum is given by
\begin{eqnarray}
    \label{eq:fish3D}
    F_{\alpha \beta} = \displaystyle \sum_{z,k,\mu\,{\rm bins}}\frac{\partial \boldsymbol{O}^{T}}{\partial \theta_\alpha} \boldsymbol{C}^{-1} \frac{\partial \boldsymbol{O}}{\partial \theta_\beta} \, ,
\end{eqnarray}
where the observable $\boldsymbol{O}$ is the observed 3D \lya forest power spectrum $P^{\rm 3D,obs}_F (\boldsymbol{k},z)$, $\boldsymbol{C}$ is the covariance matrix, and the sum is over all the bins in redshift, wavenumber and $\mu = k_\parallel/k$. Again, in principle we assume that different bins are uncorrelated in order to sum over the resulting independent Fisher matrices. 

For readers unfamiliar with \lya forecasting, the primary difference between the \lya forest Fisher matrix calculation and a galaxy counterpart, as pointed out by \cite{2014JCAP...05..023F}, is the need to evaluate the integrals over the quasar luminosity function and spectrograph noise distribution to determine the signal-to-noise level as a function of redshift, i.e.\ the \lya forest introduces the problem of accounting for the sparse distribution of quasars as a function of redshift.

The observed 3D \lya forest power spectrum can be computed from the theoretical contribution, the QLF, and the spectrograph performance, as follows \citep{2007PhRvD..76f3009M,2014JCAP...05..023F}
\begin{eqnarray}
    \label{eq:obs_P3D}
    P^{\rm 3D,obs}_F (\boldsymbol{k}) = P^{\rm 3D}_F (\boldsymbol{k})  + P^{\rm 1D}_F (k_\parallel) P^{\rm 2D}_w + P^{\rm eff}_N \, .
\end{eqnarray}
Here, the first term of rhs is the theoretical \lya forest power spectrum, the second term represents the aliasing term accounting for the 2D quasar density, and the last term is the effective noise power. Specifically, $P_w^{\rm 2D}$ is the inverse of the effective density of lines of sight that we expect given our QLF and spectrograph performance, and $P^{\rm 1D}_F$ is 1D \lya forest power spectrum, which is already measured to high signal-to-noise. In what follows the 1D \lya power spectrum will be described by the empirical fits obtained in \cite{2013A&A...559A..85P}, i.e. we will not use Eq.~(\ref{eq:P1D}) in the 3D forecast here. The reason for this is that the main contribution of the dependence with the memory of reionization in the 3D forecast is due to the reionization term in $P^{\rm 3D}_F$, so we will neglect the memory of reionization in $P^{\rm 1D}_F$ in the 3D forecast.

$P_w^{\rm 2D}$ and $P^{\rm eff}_N $ are given by
\begin{eqnarray}
    P_w^{\rm 2D} = \frac{I_2}{I_1^2 L_q} \, , \\
    P^{\rm eff}_N = \frac{I_3 l_p}{I^2_1 L_q} \, , 
\end{eqnarray}
where $L_q$ is the length of the forest in a quasar spectrum, $l_p$ is the pixel width. The $I_i$ are integrals given by
\begin{eqnarray}
    \label{eq:i1}
    I_1 &=& \int dm \frac{dn_q}{dm}w(m) \, , \\
    \label{eq:i2}
    I_2 &=& \int dm \frac{dn_q}{dm}w^2(m) \, , \\
    \label{eq:i3}
    I_3 &=& \int dm \frac{dn_q}{dm}  \sigma^2_N(m)w^2(m) \, ,
\end{eqnarray}
where $dn_q/dm$ is the luminosity function of observed quasars as a function of magnitude $m$, $\sigma_N(m)$ is the pixel noise, which is generally a function of magnitude. The weight function $w(m)$ as a function of magnitude is
\begin{eqnarray}
    w(m) = \frac{P_S/P_N(m)}{1 + P_S/P_N(m)} \, ,
\end{eqnarray}
where $P_S$ is the signal power at some typical wavenumber and $P_N(m)= \sigma^2_N(m)l_p/I_1L_q$. 

The theoretical \lya forest power spectrum \citep[e.g.][]{2003ApJ...585...34M}, which includes the memory of reionization (see MM20), 
is given by  
\begin{eqnarray}
    \label{eq:P3D}
    P^{\rm 3D}_F(\boldsymbol{k})  = b_F^2 (1 + \beta_F \mu^2)^2 P_L(k) D(\boldsymbol{k}) + 2 b_F b_\Gamma (1 + \beta_F \mu^2) P_{m,\psi} (k) \, ,
\end{eqnarray}
where the first term of rhs is the \emph{conventional} nonlinear \lya forest power spectrum with the nonlinear correction $D(k,\mu)$, and it is computed using the fits from \cite{2003ApJ...585...34M}. 
\begin{eqnarray}
    \label{eq:D}
    \ln D(k,\mu) = \left(\frac{k}{k_{\rm NL}}\right)^{\alpha_{\rm NL}} - \left(\frac{k}{k_{\rm p}}\right)^{\alpha_{\rm p}} - \left(\frac{k_\parallel}{k_{\rm v}(k)}\right)^{\alpha_{\rm v}} \, .
\end{eqnarray} 
The first term in Eq.~(\ref{eq:D}) accounts for the increase in the power spectrum due to nonlinear growth, the second term corresponds to the pressure suppression, which is subdominant for the $k$-range we are interested in for this section, and the third term corresponds to the finger-of-god effect, i.e.\  the suppression due to peculiar velocities along the line of sight. The values for the different coefficients can be found in \cite{2003ApJ...585...34M}.

The second term in the rhs of Eq.~(\ref{eq:P3D}) is the contribution due to reionization, as given by Eq.~(\ref{eq:psi}), and it is constructed using our large- and small-scale simulation results. 

Note that the aliasing term also contains information about the astrophysics of reionization, although the effect is smaller than the one in the 3D power spectrum at all redshifts. However, since Eq.~(\ref{eq:obs_P3D}) evaluates $P_F^{\rm 1D}(k_\parallel)$, lower $\mu$ bins will have an increase in significance since the memory of reionization is significantly stronger at small $k$ (or $k_\parallel$). Even though the aliasing term formally contains information regarding the astrophysics of reionization, we choose to treat it as pure noise, which is quite a conservative approach and handicaps our forecasting sensitivity, but it is also the strategy used when forecasting Baryon Acoustic Oscillations (BAO) in the 3D \lya power spectrum. 

In Figure \ref{fig:P3D_total}, we plot our observable, the observed 3D \lya power spectrum, and its components, as a function of wavenumber in the range of $0.03 \le k \le 0.70\,{\rm Mpc}^{-1}$, the $k$-bins considered in our forecast, at a given angle of $\boldsymbol{k}$ with respect to the line of sight at a low redshift $z = 2.13$, the central redshift of our lowest-redshift bin. At this redshift, the observable is dominated by ``noise'' (both aliasing term and the effective noise) at small scales. On the other hand, the 3D \lya power spectrum \emph{signal} dominates at large scales, and for smaller $k$,  the 3D \lya power spectrum becomes comparable to the ``noise'' at smaller $\mu$, i.e.\ when the power spectrum is almost perpendicular to the line of sight. The transition wavenumber coincides with the scale where the effect of the memory of reionization becomes stronger in the power spectrum. 
Both aliasing term and the effective noise are constructed using DESI QLF (with target selection) and g-band spectrograph information. Note that the aliasing term at $\mu=0.1$ is almost independent of wavenumber because $P_F^{\rm 1D}$ is almost flat at $k < 0.001\, {\rm s/km}$. The effective noise is generally independent of the wavenumber. 
Both aliasing term and the effective noise depend on the number density of quasar lines of sight and generally tend to increase with higher redshift (because of less available lines of sight). In fact, the aliasing term  starts to dominate over the signal --- even at low $k$ --- around $z \approx 3.28$, as shown in Figure \ref{fig:P3D_total_II}, which leads to a decline of the forecasting power in the higher redshifts in direct contrast to our findings in \S\ref{sec:1D}. 

\begin{figure*}
    \centering
    \includegraphics[width=\textwidth,keepaspectratio]{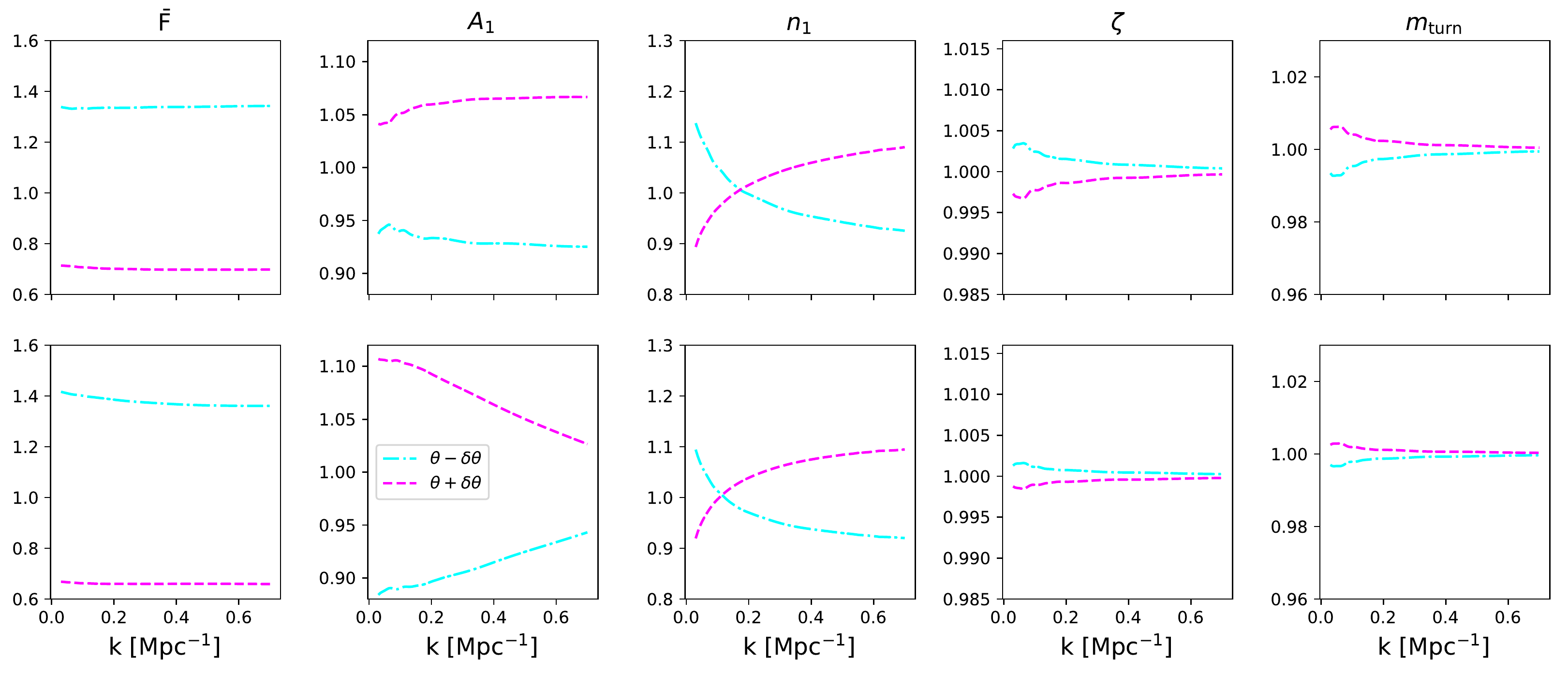}
    \caption{Ratio of the 3D \lya power spectra $P_F^{\rm 3D}(\boldsymbol{k}; \theta + \delta \theta)/P_F^{\rm 3D}(\boldsymbol{k}; \theta)$ (dashed  magenta) and $P_F^{\rm 3D}(\boldsymbol{k}; \theta - \delta \theta)/P_F^{\rm 3D}(\boldsymbol{k};\theta)$ (dot-dashed cyan), where the parameter $\theta$ is (from the left to right) the observed transmitted flux $\bar{F}$, amplitude of linear power spectrum $A_1$, tilt of the linear power spectrum $n_1$, ionization efficiency $\zeta$, and threshold mass $\mt$. Shown are the results at $z=2.13$ at $\mu = 0.1$ (top) and $0.9$ (bottom).}
    \label{fig:P3D_models}
\end{figure*}

\begin{figure*}
    \centering
    \includegraphics[width=\textwidth,keepaspectratio]{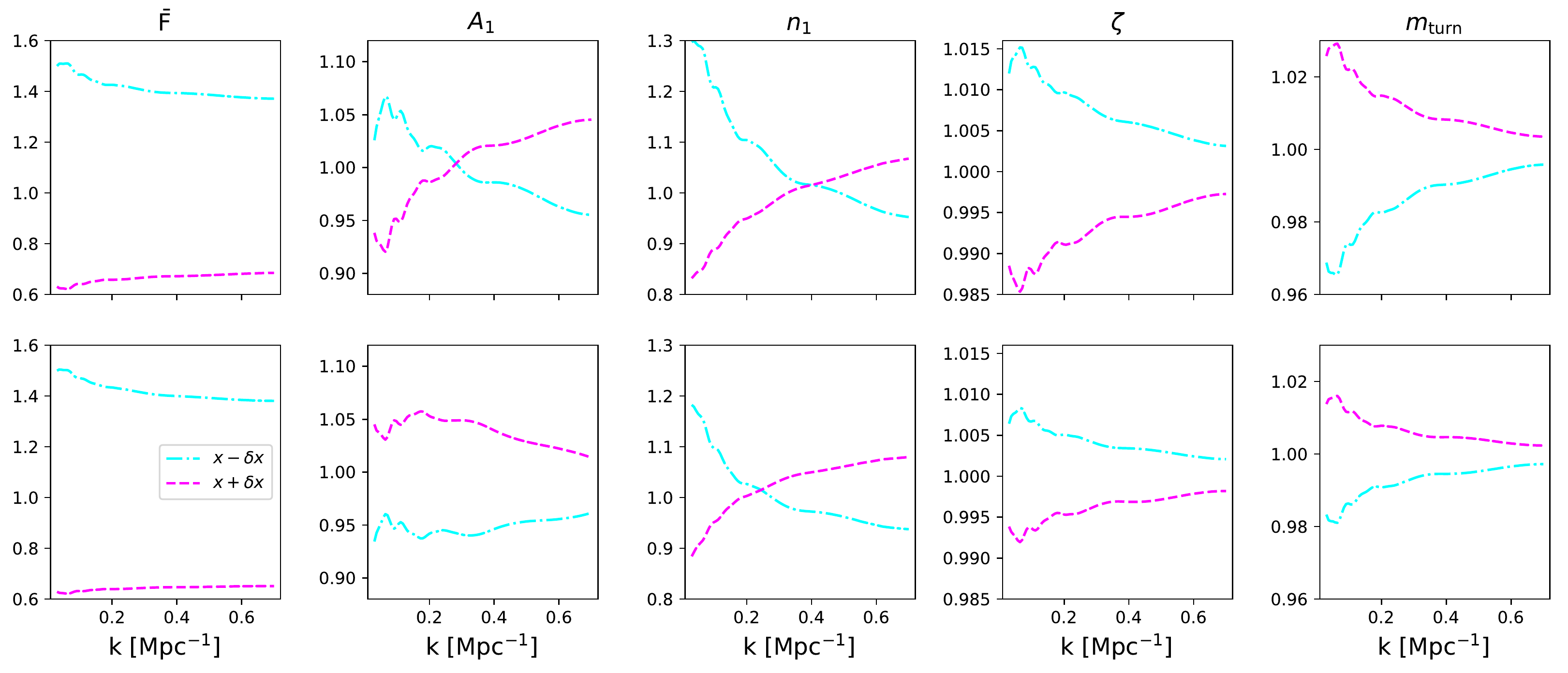}
    \caption{Same as Figure~\ref{fig:P3D_models} but for $z=3.94$.
    }
    \label{fig:P3D_models_II}
\end{figure*}

\begin{figure*}
    \centering
    \includegraphics[width=\textwidth,keepaspectratio]{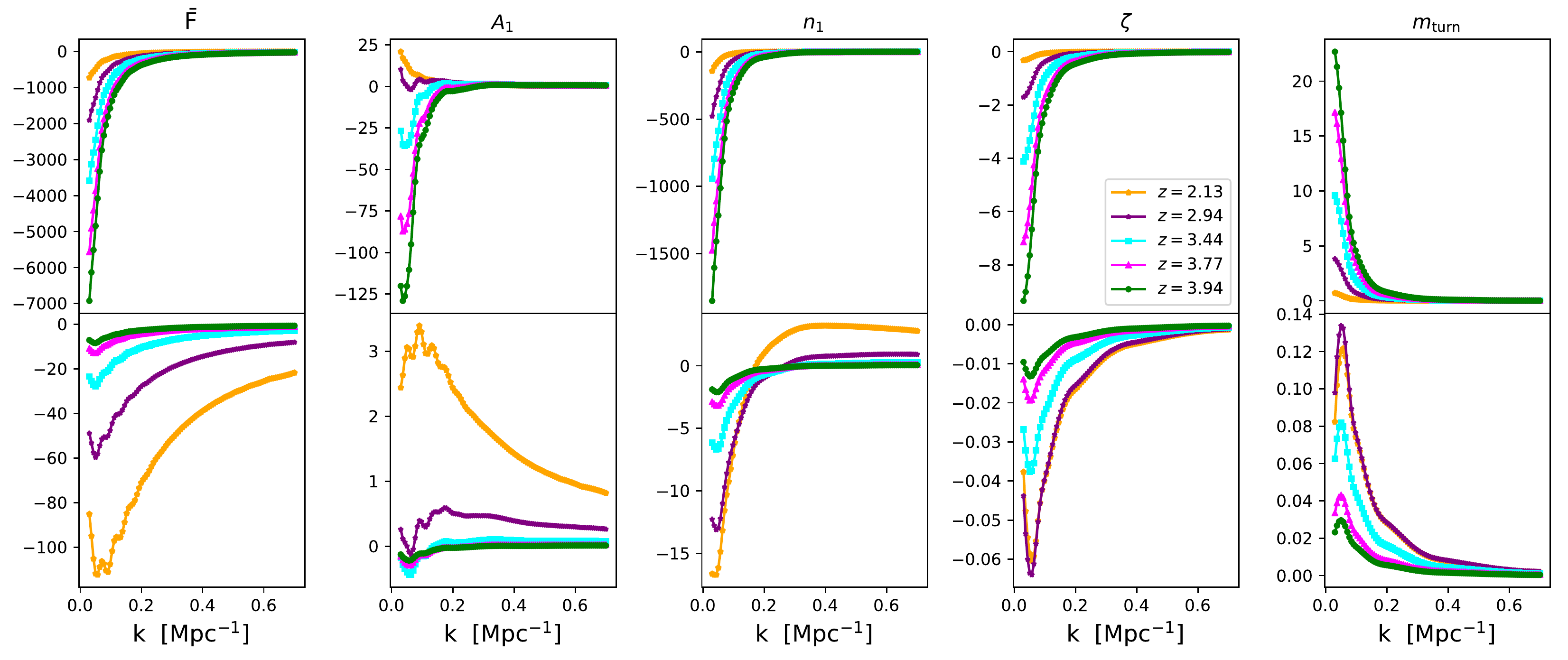}
    \caption{(Top) derivatives of the \emph{observed} 3D \lya forest power spectrum $\partial P_F^{\rm 3D, obs}/\partial \theta_\alpha$ with respect to parameter $\theta_\alpha$, where the parameter is the (from the left to right) the observed transmitted flux $\bar{F}$, amplitude of linear power spectrum $A_1$, tilt of the linear power spectrum $n_1$, ionization efficiency $\zeta$, and threshold mass $\mt$. Shown are the results as a function of wavenumber $k$ at a fixed $\mu = 0.1$, at different redshift $z=2.13/2.94/3.44/3.77/3.94$ (orange pentagons/purple stars/cyan squares/magenta triangles/green circles), respectively. 
    (Bottom) derivatives weighted by the error of the observed power spectrum, i.e.\ $(1/\Delta P)(\partial P_F^{\rm 3D, obs}/\partial \theta_\alpha)$.}
    \label{fig:partial_P3D}
\end{figure*}

\begin{figure*}
    \centering
    \includegraphics[width=\textwidth,keepaspectratio]{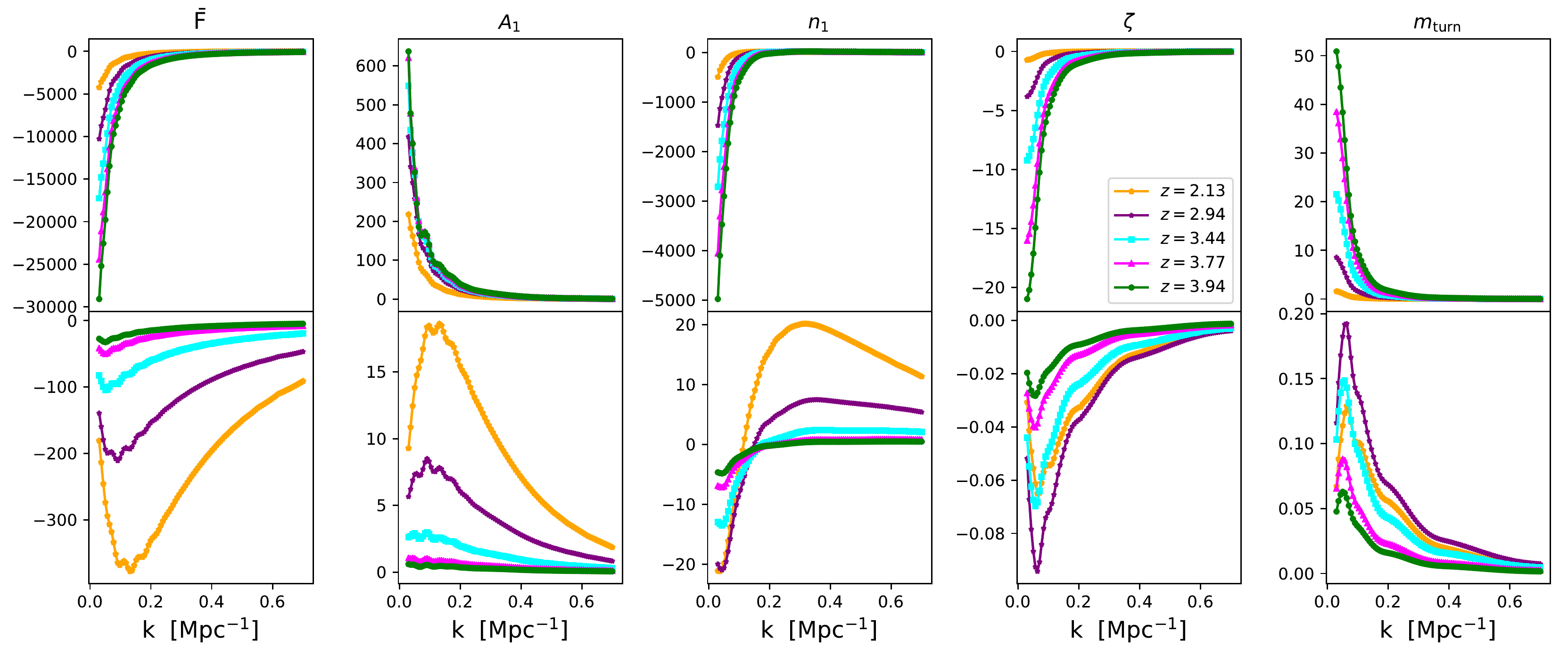}
    \caption{Same as Figure~\ref{fig:partial_P3D} but at $\mu = 0.9$.}
    \label{fig:partial_P3D_II}
\end{figure*}

In Figures~\ref{fig:P3D_models} and \ref{fig:P3D_models_II}, we illustrate the difference of the theoretical 3D \lya power spectra at $z=2.13$ and $3.94$ respectively, when the forecast parameters are varied around their fiducial values, which is irrelevant of the specific instrument used for measurements.  In general, we find that large scales and high redshifts present the best conditions to extract the astrophysics of reionization. Similar to Figure \ref{fig:partial_P1D}, we see that the nuisance parameters are again the ones playing the dominant roles in the forecast. This is the expected behavior since the memory of reionization does not surpass $50$ per cent of the expected theoretical signal for a model that considers realistic models of reionization and HEMD gas (see MM20).

At low redshifts, a smaller observed transmitted flux leads to an increase in $P_F^{\rm 3D}$, since $b_F^2 (\bar{F} + \delta \bar{F}) = 0.0126$ vs $b_F^2 (\bar{F} - \delta \bar{F}) = 0.0235$. At higher redshifts, reionization increases the separation between $\bar{F}$ models mainly due to the fact that the lesser forest (larger mean flux) has a slower change of the transparency of the IGM. In addition, the radiation bias needed to match observations for the lower flux model is significantly larger than for the model with the larger flux. 

For $A_1$, we see the expected increase in flux power that comes with the increase of mass power, particularly closer to the line of sight. For more transverse directions, the stronger increase on $P_F^{\rm 3D}$ gets suppressed because of peculiar velocities at larger wavenumbers. At high $z$, the reionization term is stronger and we see that increasing $A_1$ can actually lead to a smaller flux power at large scales. This is due to a smaller mass power that leads to fewer structures, and hence less HEMD gas, which results in a faster decrease in the change of transparency of the IGM. Also, a larger radiation bias is needed to match the mean transmitted flux. Besides, reionization happens earlier for a larger amplitude of the linear matter power spectrum, since more star-forming galaxies form early, which translates into a smaller memory of reionization in the \lya forest. Parallel directions suppress this behavior due to the extra factor of $(1 + \beta_F \mu^2)$ in the first term of Eq.~(\ref{eq:P3D}).

Meanwhile, at low redshift a decrease in $n_1$ leads to a larger flux power at small $k$ because of the isotropic increase in power due to nonlinear growth, which is enhanced for the $n_1 - \delta n_1$ scenario ($k_{\rm NL} = 5.07$ $h^{-1}$ Mpc$^{-1}$ and $\alpha_{\rm NL} = 0.579$) compared to the $n_1 + \delta n_1$ model ($k_{\rm NL} = 8.30$ $h^{-1}$ Mpc$^{-1}$ and $\alpha_{\rm NL} = 0.551$). At large $k$, the initially larger matter power spectrum for $n_1 - \delta n_1$ and suppression due to peculiar velocities leads to larger flux power. At larger redshifts, reionization perturbs the isotropy present at lower $z$, especially at low $k$ and $\mu$ where the memory of reionization is stronger. We see that the enhanced effect at small $\mu$ of the memory of reionization moves the crossing-point to smaller scales. Besides, we see a significant increase of the power at large scales for smaller tilt fueled both by nonlinear growth and a delayed reionization; however, due to the extra factor of $(1 + \beta_F \mu^2)^2$, along the line of sight we see that this enhancement is diminished. 



Figures~\ref{fig:P3D_models} and \ref{fig:P3D_models_II} generally show that delayed reionization leads to a larger flux power spectrum. Three regimes lead to a stronger effect, in general --- (i) at high redshifts, when the IGM has the least amount of time to relax the additional energy injected by the reionization process, (ii) at small wavenumber, since the cross-correlation between matter and neutral hydrogen fraction couples to the ionized bubble size, and (iii) in the directions transverse to the line of sight, particularly for the variation of reionization parameters, due to the Kaiser effect, because the 3D \lya power spectrum depends on reionization parameters through $P_{m,\psi}$ in the second term of the rhs of Eq.~(\ref{eq:P3D}), while the power spectrum itself is dominated by the first term. Therefore $(\partial P_F^{\rm 3D}/\partial \theta)/P_F^{\rm 3D} \appropto 1/(1+\beta_F\mu^2)$. 
The ``weird'' behavior at the smallest wavenumbers is likely due to the numerical error arising from the small number of modes per $k$-bin in our $P_{m,x_{\rm HI}}$. 
Based only on the ratios shown in this figure, we find that the sweet spot for constraining the astrophysics of reionization is, again, at high redshifts, in the large scales, and in the transverse direction with respect to the line of sight.
Nonetheless, the derivatives of the observed 3D power spectrum, and therefore of the 3D \lya power spectrum, are actually larger for the directions along the line of sight. This is due to the raw increase of power along the line of sight, as can be seen in Figure \ref{fig:P3D_total}. Similar to what happened for the 1D derivatives, even the weighted derivatives will not preserve this sweet spot due to the strength of the aliasing term, which will dominate the signal at high $z$. 


In Figures~\ref{fig:partial_P3D} and \ref{fig:partial_P3D_II}, we plot the derivatives, and those weighted by the error (see Eq.~\ref{eq:cov} below), of the observed 3D \lya power spectrum with respect to the forecast parameters at the fixed 
$\mu=0.1$ and 0.9, respectively. Similar to the 1D case, the derivatives are larger (in absolute value) for higher redshifts, larger scales and in the longitudinal directions, due to the increase in flux power. However, weighting by the fiducial covariance disrupts these features primarily because of the drop in the effective 2D density of quasar lines of sight. As a consequence, the lower redshift bins affect the forecast the most. 

\begin{figure*}
    \centering
    \includegraphics[width=\textwidth,keepaspectratio]{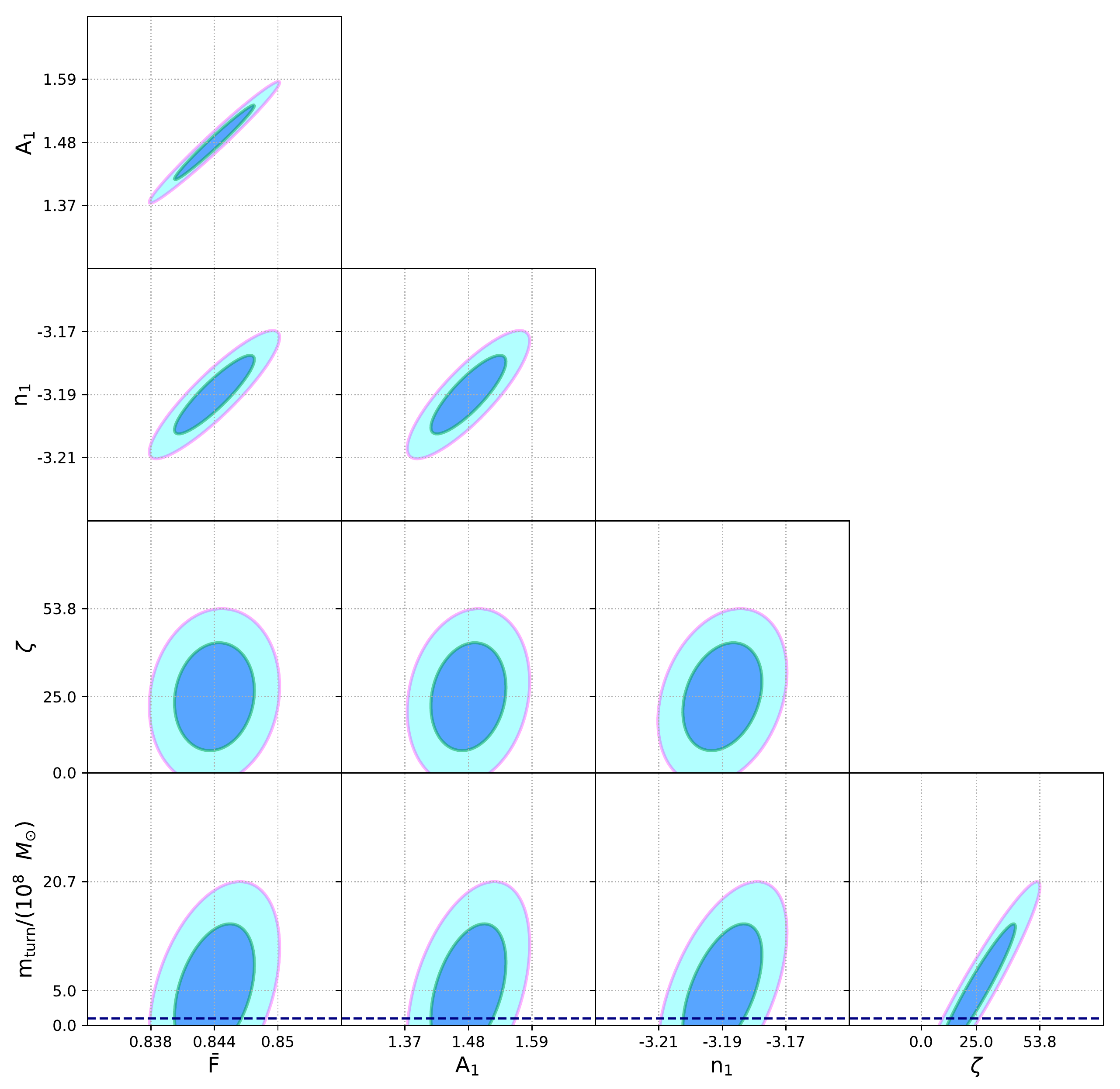}
    \caption{Forecast of parameter estimation from the 3D \lya forest power spectrum using DESI specs: DESI QLF and DESI spectrograph performance, and accounting for all redshift bins, due to the effect of memory of reionization (including the effect due to patchiness and the small-scale HEMD gas). The DESI specs are used to compute Eqs.~(\ref{eq:i1} -- \ref{eq:i3}), which are needed to construct the effective number density of quasar lines of sight and the effective noise power. All contours are centered on our fiducial values, and the blue (cyan) ellipses correspond to the $1\sigma$ ($2\sigma$) contours. We opt to put a physical lower limit for $\mt = 10^8 \,{\rm M}_\odot$ (horizontal blue dashed line), corresponding to efficient atomic cooling of star-forming galaxies.  
    }
    \label{fig:DESI_forecast}
\end{figure*}

Furthermore, a direct comparison with Figure \ref{fig:partial_P1D} reveals that the derivatives with respect to mean transmitted flux and reionization parameters follow the same trends, in contrast with the discrepancy present for the derivatives with respect to cosmological parameters. The discrepancy in the high redshift results for $A_1$ in the transverse directions is due to the behavior shown in Figures~\ref{fig:P3D_models} and \ref{fig:P3D_models_II}, which originates from the role of $A_1$ in the memory of reionization. Analogously, for the case of $n_1$, we see that the derivatives are negative and sharp at large scales, consistent with the behaviour seen in Figures~\ref{fig:P3D_models} and \ref{fig:P3D_models_II}. Note that the sign of the derivatives is a consequence of the competition between (i) matter power, which increases at large scales with smaller $n_1$, (ii) nonlinear growth, which is enhanced for smaller $n_1$, (iii) timing of reionization, which is delayed for small $n_1$, and (iv) transparency of the IGM, which changes more rapidly for larger $n_1$ due to fewer structures present in our high-resolution simulations and a larger radiation bias required to match observations. We see that the derivatives are particularly enhanced for directions along the line of sight due to the accompanying increase in signal.  


The last ingredient needed for our forecast is the covariance matrix describing the band power errors for each $z,k$, and $\mu$ bins. For a given survey volume V and redshift bin $i$, the covariance matrix for the mode $\boldsymbol{k}$ is given by 
\begin{eqnarray}
    \label{eq:cov}
    C_{ij\boldsymbol{k}\boldsymbol{k}'} = \delta_{ij}\, \delta_{\boldsymbol{k},\boldsymbol{k}'}\,\langle \Delta P^2 \rangle  = \delta_{ij}\, \delta_{\boldsymbol{k},\boldsymbol{k}'}\,\frac{4 \pi^2}{V k^2 \Delta k \Delta \mu} \left({P^{\rm 3D,obs}_F}(z_i, \boldsymbol{k})\right)^2 \, ,
\end{eqnarray}
where $\Delta k$ and $\Delta \mu$ are the respective bin widths, and the bins are assumed small enough that the Fisher matrix calculation is an integral over $k$ and $\mu$. To compute the total covariance matrix, we assume that the errors are uncorrelated for different redshifts, and thus the Fisher matrix can be summed over the redshift bins.

\subsection{Forecast with DESI}
\label{ssec:DESI}

Now we show the centerpiece of this work, a first forecast for the thermal relics of reionization -- and HEMD gas -- in the DESI 3D \lya forest power spectrum. 


We show the forecast capability of DESI to constrain the astrophysics of reionization using all redshift bins considered in this work in Figure~\ref{fig:DESI_forecast}. As expected, the 3D forecast is significantly more capable of constraining the nuisance parameters than its 1D counterpart. For example, we obtain $\Delta A_1 = 0.04$ and $\Delta n_1 = 0.008$
for the 3D forecast compared to $\Delta A_1 = 0.06$ ($0.05$) and $\Delta n_1 = 0.06$ ($0.03$)
for the High-z (Low-z) 1D forecast.  Note that we have recovered the expected degeneracy between the ionization efficiency and the threshold mass, but the angle of the ellipse is given by $\theta_{\rm 3D} = 28.0^{\circ}$, which is almost the degeneracy angle recovered by our hypothetical Low-z survey in \S\ref{ssec:eBOSS}. This is perhaps not surprising since the higher redshifts are weighted unfavorably due to the increase in noise level (see Figure \ref{fig:P3D_total}), but it is also an evidence that the lower redshifts are probably less sensitive to the midpoint of reionization due to a lower ability to constrain the ionization efficiency. 
Overall the similarities between our 1D and 3D forecasts are a sign of how optimistic our ideal 1D forecast is, or, alternatively, how pessimistic our conservative scenario for the 3D power spectrum really is. 

For reionization parameters, we highlight that our forecast errors\footnote{While the value for $\mt$ seems to be compatible with zero because $\mt = (5.00 \pm 6.34) \times 10^8 $ M$_\odot$ in Figure~\ref{fig:DESI_forecast}, here we opt to use a physical lower limit on $\mt$ based on the efficiency of atomic cooling, $\mt = 10^8$ M$_\odot$, corresponding to efficient atomic cooling of star-forming galaxies \citep{2019MNRAS.484..933P}.
}, $\zeta = 25.0 \pm 11.6$ and $\log_{10}(\mt / $M$_\odot) = 8.7 ^{+0.36}_{-0.70}$, are surprisingly competitive with previously forecast errors on the same parameters using the 21~cm power spectrum measurement with a mock 1000h observation with the Hydrogen Epoch of Reionization Array (HERA), $\zeta = 34.69^{+9.48}_{-6.50}$ \citep{2017MNRAS.472.2651G} and $\log_{10}(\mt / $M$_\odot) = 8.80^{+0.27}_{-0.26}$ \citep{2019MNRAS.484..933P}. 
The forecast errors from the \lya power spectrum with DESI are  slightly larger than those from the 21~cm power spectrum with HERA; specifically, $\sigma_{\rm DESI} / \sigma_{\rm HERA} \approx 1.5$ for $\zeta$ and $\sigma_{\rm DESI}/\sigma_{\rm HERA} \approx 2.0$ for $\log_{10}(\mt / $M$_\odot)$. On the other hand, the constraint from the current EoR observations combined with model-dependent priors from high-z galaxy observations (e.g., observations of the faint end of the galaxy luminosity function) is $\zeta = 28^{+52}_{-18}$  \citep{2017MNRAS.465.4838G}. Thus our forecast constraint with DESI is about three times tighter than the current EoR constraint, i.e.~$\sigma_{\rm DESI} / \sigma_{\rm current} \approx 1/3$ for $\zeta$. From this corner plot, we conclude that DESI will likely be able to constrain the astrophysics of reionization even in the case where we have neglected all information from the aliasing term, if many observational and theoretical challenges are tackled in order to make the 3D \lya forest power spectrum measurement come true. However, even if systematics make a potential measurement of the 3D \lya forest power spectrum challenging, constraining the astrophysics of reionization using only the 1D \lya forest power spectrum measured by DESI should be achievable. 

Given the constraints of the astrophysics of reionization with DESI, we now show the implications of our results to the global reionization history, 
not only for its intrinsic value, but also because the constraints on the EoR parameters are sensitive to the accuracy of our parametrization of the astrophysics of reionization and to the assumptions made about the sources of UV photos that reionize the IGM, in addition to other systematic uncertainties introduced due to the way we model the EoR \citep[e.g.][]{2021MNRAS.504.1555M}. However, constraints on the global history of reionization are more robust against these parametrizations of the physics of reionization \citep{2017MNRAS.465.4838G}. 


\begin{figure}
    \centering
    \includegraphics[width=\columnwidth]{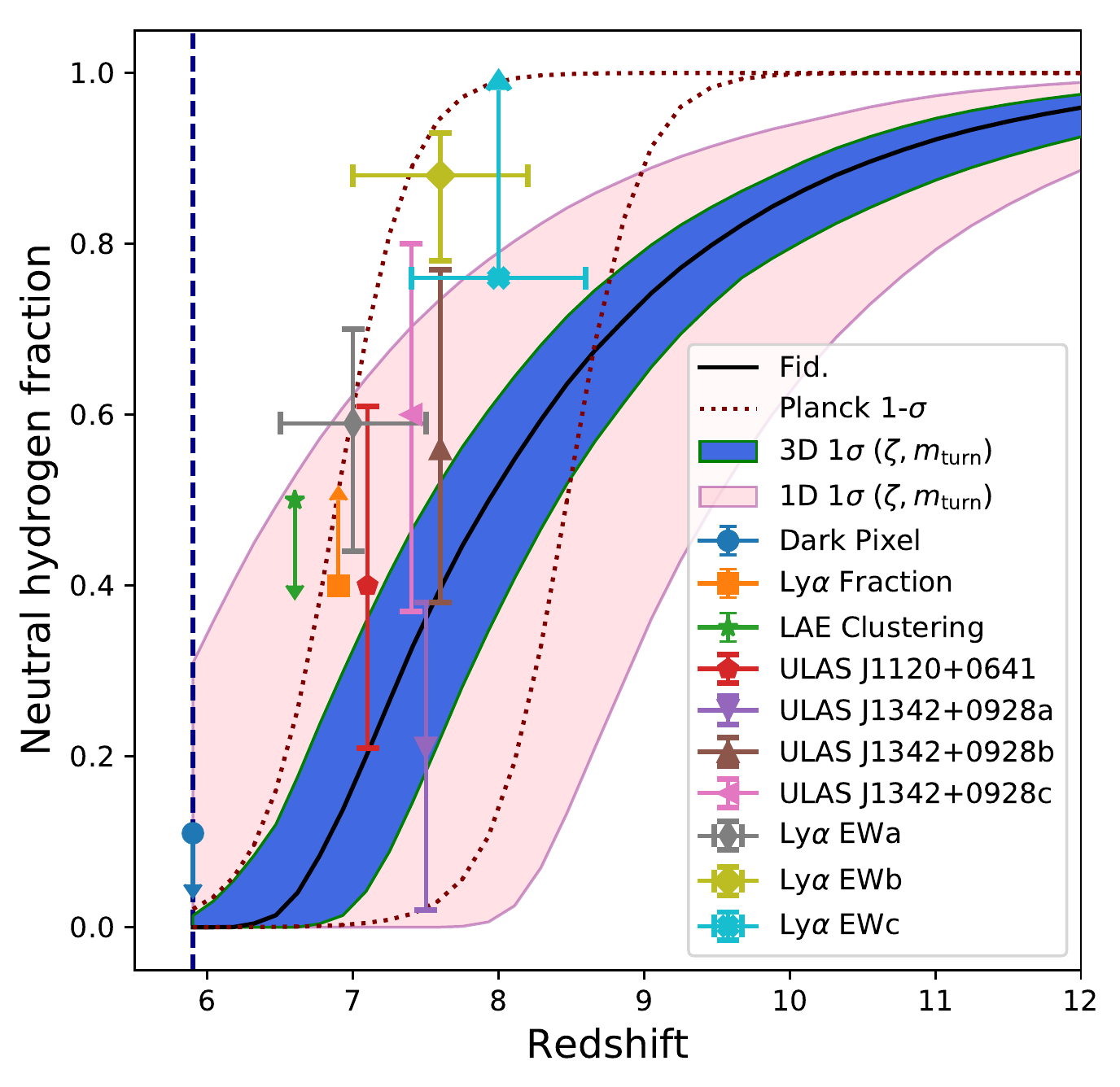}
    \caption{Implication for the global reionization history. Shown is the redshift evolution of the volume-weighted neutral hydrogen fraction of the IGM for our fiducial model (black solid line), and the inferred region corresponding to a selection of mixed models allowed by our 1$\sigma$ contour on $\zeta$ and $\mt$ from the 3D \lya forest power spectrum with DESI (blue region), and from the 1D High-z survey (pink region).
    We also show the constraints due to dark pixel fraction (blue circles)\citep{2015MNRAS.447..499M}, \lya fraction evolution (orange squares)\citep{2015MNRAS.446..566M}, clustering of \lya emitting galaxies (green stars)\citep{2010ApJ...723..869O,2015MNRAS.453.1843S}, QSO damping wings (red pentagons, purple/brown/pink triangles) \citep{2017MNRAS.466.4239G,2019MNRAS.484.5094G,2018Natur.553..473B,2018ApJ...864..142D}, \lya equivalent width distribution (grey diamonds/yellow squares/cyan crosses)\citep{2018ApJ...856....2M,2019ApJ...878...12H,2019MNRAS.485.3947M},
    and the inferred constraint from Planck using the ``TANH'' model (dotted maroon points)\citep{2020A&A...641A...6P}. In order to improve the clarity of the figure, we move some of the clustered constraints by $\delta z= 0.1$. Our lower limit for reionization is taken at $z = 5.90$ (dashed navy blue vertical line).}
    \label{fig:global}
\end{figure}

In Figure \ref{fig:global}, we illustrate the ability of DESI to constrain the global history of reionization using the 3D and 1D \lya forest power spectrum. We also include constraints based on the observations of the \lya and Ly$\beta$ forest dark fraction \citep{2015MNRAS.447..499M}, \lya emission from galaxies \citep{2015MNRAS.446..566M}, clustering of \lya emitters \citep{2010ApJ...723..869O,2015MNRAS.453.1843S}, the damping wings of ULAS J1120+0641 \citep{2017MNRAS.466.4239G} and ULAS J1342+0928 \citep{2019MNRAS.484.5094G,2018Natur.553..473B,2018ApJ...864..142D}, and the redshift evolution of the \lya equivalent width distribution in Lyman-break galaxies \citep{2018ApJ...856....2M,2019ApJ...878...12H,2019MNRAS.485.3947M}. To infer a constraint on the evolution of neutral hydrogen with the memory of reionization, we plot the inferred constraint corresponding to the 1$\sigma$ region in the $\zeta$-$\mt$ parameter space marginalized over other nuisance parameters. 

With the assumption of our fiducial model, our results from the 3D \lya forest power spectrum with DESI are translated to the constraint of the midpoint of reionization (defined as the redshift when global neutral fraction is 0.5) between  $z_{\rm mid}=7.52$ and $z_{\rm mid}=8.41$, and the duration of reionization (defined as the redshift interval between global neutral fraction at 0.25 and 0.75)
between $\Delta z = 1.52$ and $\Delta z = 2.62$. Meanwhile, for the 1D High-z survey introduced in \S\ref{sec:1D}, we find $6.50 < z_{\rm mid} < 9.49$ and $1.21 < \Delta z < 2.73$.\footnote{For the 1D inferred constraint, we additionally impose a lower limit on $\zeta \geq 10$ (as done in \citealt{2017MNRAS.472.2651G}), since some reionization models allowed by our $1\sigma$ region exhibit unnaturally late reionization, e.g. midpoint of reionization at $z_{\rm mid}< 6.0$, which would be in large tension with current observations \citep[e.g $z_{\rm mid} = 7.69 \pm 0.79$; ][]{2020A&A...641A...6P}.} 
The ranges for the midpoint of reionization are consistent with $z_{\rm mid} = 7.7$ inferred by \cite{2020A&A...641A...6P} using the TANH model. Our results of the duration of reionization are also consistent with the current observational indication of reionization occurring late and fast.

We close this section by reminding the readers that, due to the limitations of our analysis, the interesting results are not the central values, because they are set by the fiducial model, but the error forecasts. Besides, DESI might be able to improve the constraints by performing joint analysis with the 1D \lya power spectrum and using the cross-correlation of quasars with the \lya forest \citep[e.g. ][]{2013JCAP...05..018F}, or even by taking advantage of external information such as using overdense regions \citep[e.g. ][]{2017ApJ...839..131C} at the relevant redshifts  to cross-correlate with intensity mapping of \lya emission \citep[e.g. ][]{2018MNRAS.481.1320C}, or combining with external data sets like GRB afterglows \citep{2021arXiv210502293L}. Furthermore, using the dark pixel fraction to limit the models for the tail-end of reionization will enhance the forecasting strength. Even using only the 3D \lya forest power spectrum will probably outperform our forecast since we have forsaken the signal present in the aliasing term (as exemplified in \S\ref{sec:disc}). In future work, we will explore these exciting possibilities.  

\section{Discussion}
\label{sec:disc}

\subsection{Forecasting strategy} 

The forecasting strategy in this paper ignores the signal from the aliasing term, which is the same strategy as used in forecasting BAO in the 3D \lya forest\footnote{Patrick McDonald (private communication).}. Our reasoning for this choice was to select the most conservative approach. However, an alternative strategy for the 3D forecast consists on allowing for variations in $P_F^{\rm 1D}$ while restricting the number of modes available to $k_\perp$ smaller or equal to the Nyquist frequency corresponding to the mean separation of quasar lines of sight \citep[e.g.][]{2014JCAP...05..023F}. This is the approach utilized when forecasting the sensitivity to neutrino masses in the 3D \lya forest power spectrum. 

We highlight that both methods correspond to different prescriptions to avoid double counting of information, i.e.\ avoid the scenario where we have more modes than data points. Due to the nature of our signal, the two strategies produce different results. As an example of the impact of this choice, we tabulate the loss of constraining power for our first and central redshift bins utilizing the mean separations reported in Table 1 of \cite{2018MNRAS.477.1814C} as our assigned Nyquist frequencies in Table \ref{tab:loss}. For comparison purposes, we also use the BOSS and eBOSS mean separations. We label the implementations with different Nyquist frequencies as {\sc INSTRUMENT-Ny}, while the strategy used throughout this paper is label as ``{\sc DESI-nop}''. Furthermore, in order to avoid extrapolating our simulation results for the 1D \lya power spectrum, we restrict our analysis to $0.1 \le k \le 0.7\, {\rm Mpc}^{-1}$ for the strategies that implement a cutoff and allow $P^{\rm 1D}$ to vary\footnote{We remind the readers that in Eq.~(\ref{eq:obs_P3D}) the 1D power spectrum is evaluated at $k_\parallel$, so we make this change in wavenumbers such that the first $\mu$-bin is almost entirely extrapolation-free (our {\sc 21cmFAST} boxes have a side of $L = 400$ Mpc, so our first mode is approximately 0.016 Mpc$^{-1}$). Therefore, we cannot make a completely direct comparison between the strategy that utilizes the aliasing term as a signal with the forecast that treats the aliasing term as pure noise.}. 
Due to the change in $k$-bins, we only compare the change in the forecast errors for the astrophysics of reionization.   

As illustrated in Table \ref{tab:loss}, the loss of forecasting strength is minimal at low redshifts. Nevertheless, the effect is considerably large at high redshifts, where we see the constraints of the astrophysics of reionization using {\sc DESI-Ny} are roughly a factor of two as tight as using {\sc DESI-nop}, because {\sc DESI-nop} treats the aliasing term as pure noise. Although {\sc DESI-nop} effectively misses the sweet spot to constrain reionization, we opted to use {\sc DESI-nop} in \S\ref{ssec:DESI} to forecast the ability of DESI to constrain the astrophysics of reionization in a conservative fashion. This choice is motivated due to the uncertainties in our modeling of the \lya forest (see \S\ref{ssec:cav}) and because of the larger box needed to compute the aliasing signal -- with memory of reionization -- at low $\mu$. 


\begin{table}
    \centering
    \caption{Estimate of the impact of 3D \lya forecasting strategy. {\sc BOSS-Ny}, eBOSS-{\sc Ny}, and {\sc DESI-Ny} allow for the aliasing term to depend on the parameters of the forecast and have a cutoff based on the mean separation between quasar lines of sight. In contrast, {\sc DESI-nop} has no cutoff but treats the aliasing term as pure noise. The mean separations are obtained from \citet{2018MNRAS.477.1814C}.}
    \begin{tabular}{ccccc}
    \hline \hline
 {} &   Strategy  & Mean separation &  \multicolumn{2}{c}{Forecast errors}  \\
 {} &   {} & [arcmin]  &  $\Delta\zeta$ &  $\Delta\mt$ [$\times 10^8 M_\odot$]  \\
    \hline
\multirow{4}{*}{\rotatebox[origin=c]{90}{$z = 2.13$}} &   {\sc BOSS-Ny} & 15  & 83.559 & 42.535  \\
{} &    eBOSS-{\sc Ny}     & 10   &  57.537  & 27.599   \\
{} &    {\sc DESI-Ny}      & 8.1  &  55.296 & 26.237  \\
{} &    {\sc DESI-nop}     & ---  & 48.322 & 29.228    \\
    \hline
\multirow{4}{*}{\rotatebox[origin=c]{90}{$z = 2.95$}} &    {\sc BOSS-Ny}  & 15      & 41.710 & 17.631  \\
{} &    eBOSS-{\sc Ny} & 10     & 33.618 & 13.382  \\
{} &    {\sc DESI-Ny}  & 8.1    & 32.647 & 13.028  \\
{} &    {\sc DESI-nop} & ---     & 45.598 & 30.825  \\
    \hline \hline 
    \end{tabular}
    \label{tab:loss}
\end{table}

\subsection{Caveats}
\label{ssec:cav}
In \S\ref{ssec:DESI} we demonstrate the power of DESI that converts the memory of reionization from a broadband systematic to an emergent window of the epoch of reionization. 
However, due to the limitations in our account of various physics and systematics, 
the values presented throughout this paper are intended to give a good estimate of what we can expect from DESI -- and DESI-generation surveys -- regarding the extraction of the astrophysics of reionization from the \lya forest, instead of claiming it as a complete formalism or precision cosmology yet. 
The approach utilized in this work has multiple areas for improvements, e.g.\ on modeling the IGM, cosmic reionization, the \lya forest, and the thermal evolution of the IGM. 

With respect to the caveats and limitations of our hybrid methodology to compute the memory of reionization in the \lya forest, Section 4.3 of MM20 has already explicitly mentioned and quantified the main sources of uncertainty, e.g. absence of \ion{He}{II} reionization, clustering of ionizing sources, and AGN feedback. Of these effects the main concern in the context of this forecast is the loss of thermal sensitivity to thermal relics of \ion{H}{I} reionization due to \ion{He}{II} reionization. Future work will quantify the impact of this effect on our results. We speculate that the inclusion of \ion{He}{II} will lead to an overall increase of forecasting strength, because in that case the signal in the aliasing noise can be included, once \ion{He}{II} reionization has been tackled. 
In addition, this effect should not be treated as \emph{only} a broadband systematic since the thermal relics from \ion{He}{II} reionization may become a novel window into the last phase transition of the Universe.   

Conventionally, the \lya forest power spectrum is computed from numerical simulations with large enough box sizes, say, 80\,cMpc on each side, to reach the convergence of the statistics. In our work, however, the base \lya forest power spectra are modelled analytically, following \cite{2003ApJ...585...34M}, and our hybrid numerical simulations are only used to estimate the impact of the memory of reionization in the \lya forest. Specifically, the small-scale high-resolution simulations are employed to compute the transparency of the IGM, and the large-scale low-resolution simulations are used to obtain the cross-correlation between matter and neutral fraction field for the duration of reionization. As such, while the small box size in the former does not affect the convergence of \emph{base} \lya statistics, it might introduce errors to the estimation of the IGM transparency. For example, if the box size is reduced from 2551 ckpc to 1275 ckpc, the IGM transparency is changed by $< 10 \%$ \citep[see \S5.3 of][]{2018MNRAS.474.2173H}. Further convergence tests of the box size may be required for improving the accuracy of forecasts in future work.


Improvements in the simulations in \cite{2003ApJ...585...34M} that are the cornerstone of \lya forecasts can provide the important avenues for improvements of future forecasts aiming at the memory of reionization. \cite{2015JCAP...12..017A} already improved along this direction with results tabulated up to $z=3$; however, further improvement to even higher redshifts would be particularly beneficial for future work. In particular, for our work these simulations are responsible for the nuisance parameters, bias factors, and modeling of the nonlinear corrections, while our hybrid simulations include the role of the nuisance parameters in the memory of reionization. The main disadvantage of our methodology is that the results from \cite{2003ApJ...585...34M} were tabulated at low redshifts ($z = 2.25$). Therefore one must extrapolate to higher redshifts where the validity of the results introduces another source of systematic error. Thus future work will include the theoretical model of the non-linear corrections based on \cite{2015JCAP...12..017A}. Moreover, the inclusion of ``new'' physics, like HEMD gas, provides compelling reasons to update these simulations.


Throughout this work we have parametrized the astrophysics of reionization with only two parameters ($\zeta$ and $\mt$), both are constant even though dependence with halo mass and redshifts should be expected. More sophisticated models \citep[e.g. ][]{2019MNRAS.484..933P} utilize four parameters to allow for these dependencies. Based on the results presented in MM20, we have not considered changes on the heating physics of the cosmic dawn, which is justified for the large scale part of our computation; however, there is still the open question of the role of X-ray preheating on the HEMD gas since X-rays can wipe out the small-scale structure before reionization, leading to a lesser effect on the \lya forest. In addition, we have chosen to not change the mean free path of ionizing photons due to its smaller role in the memory of reionization for values not exceedingly small. In addition to these active choices, other passive choices/assumptions in \texttt{21cmFAST}, like the choice of halo mass function (HMF) \citep{2021MNRAS.504.1555M} and the use of idealized halo histories instead of merger tress \citep{2020arXiv201209189M}, can lead to more sources of systematic uncertainties. Future work will address how allowing for redshift and halo mass dependence of $\zeta$ and $\mt$ analogues and the choice of HMF influence the results of our forecast. Likewise, future efforts will be directed to quantify the role of X-ray preheating in the HEMD phase using robust X-ray models.         

\subsection{``Price'' for extracting the astrophysics of reionization}

Naively, the inclusion of the astrophysics of reionization into the forecast implies a lower performance when obtaining cosmological information from $P^{\rm 3D}_F$. 
Finally, we address the question of \emph{What is the price for extracting the astrophysics of reionization with the \lya forest?} Before answering we must clarify that realistically there is no ``price'' since ignoring the memory of reionization can bias all Bayesian inference of cosmological information due to the strong effect it can have, particularly for the 3D power spectrum. Nonetheless, the question has value as a quantitative statement for the extent the ability to obtain cosmological information is damped due to having to constrain these astrophysics simultaneously. In order to quantify this performance, we compute our Fisher matrix and compare the forecast errors we have computed in \S\ref{ssec:DESI} to the ones obtained for a scenario where we turn off all reionization and HEMD phase of the temperature-density relation of the IGM.
We see a reduction on the forecast errors for the nuisance parameters of roughly a factor of two in comparison to the errors with reionization. In addition to artificially decreasing the error bars, cosmological parameters might be biased if reionization parameters are not taken into account. Thus this value should not be interpreted as a statement of an increase in the eventual inferred errors on the non-reionization parameters of the forecast, but instead as a sign that future work is needed to address the different systematic uncertainties in this forecast.

\section{Summary and Conclusions}
\label{sec:conc}

DESI was originally designed, among other science goals, to probe the post-reionization epoch using the \lya forest power spectrum, i.e.\ to measure a plethora of quasar spectra to trace the underlying matter content primarily between $2<z<4$. In MM20, we demonstrated that DESI has the potential to transform the broadband systematic signal due to the memory of reionization in the \lya forest into a window to the epoch of reionization. In this paper, we forecast, for the first time, the capability of DESI, and in general DESI-generation \lya surveys, to constrain the astrophysics of reionization through the impact of patchy reionization and HEMD gas on the \lya forest power spectra. 

Using DESI QLF and g-band spectrograph performance, we show the expected constraining power of DESI, with only the contribution due to 3D \lya forest power spectrum. Our Fisher forecast estimates the $1\sigma$ error on the ionization efficiency as $\zeta = 25 \pm 11.6$, and that on the mass threshold of haloes that host star-forming galaxies as $\log_{10}(\mt / $M$_\odot) = 8.7 ^{+0.36}_{-0.70}$, by marginalizing over the amplitude and tilt of the linear matter power spectrum and the mean observed transmitted \lya flux. Here the central values are given by our fiducial model and the lower limit for $\mt$ is imposed by the assumption that efficient atomic cooling is the dominant mechanism for star-forming. These $1\sigma$ errors are surprisingly competitive with -- but still a little larger than -- previously forecast errors using the 21~cm power spectrum mock observations of HERA with 1000h integration time \citep{2017MNRAS.472.2651G,2019MNRAS.484..933P}, 
specifically, $\sigma_{\rm DESI} / \sigma_{\rm HERA} \approx 1.5$ for $\zeta$ and $\sigma_{\rm DESI}/\sigma_{\rm HERA} \approx 2.0$ for $\log_{10}(\mt / $M$_\odot)$. Besides, our forecast constraint with DESI is about three times tighter than the constraint from the current EoR observations combined with model-dependent priors from high-z galaxy observations \citep{2017MNRAS.465.4838G}, i.e.~$\sigma_{\rm DESI} / \sigma_{\rm current} \approx 1/3$ for $\zeta$.

This forecast capability of DESI on the constraints of reionization parameters can infer the global reionization history. We illustrate that DESI may constrain the midpoint of reionization and duration of reionization in the $1\sigma$ range of $7.52 < z_{\rm re} < 8.41$ and $1.52 < \Delta z < 2.62$, respectively, where the central values in the range are given by our fiducial, late and fast reionization model that is consistent with the current state-of-the-art observations. Likewise, we forecast the capability of an eBOSS-like High-z survey to constrain the midpoint and duration of reionization using only the 1D \lya power spectrum in the $1\sigma$ range of $6.50 < z_{\rm mid} < 9.49$ and $1.21 < \Delta z < 2.73$.

In principle, 1D \lya forest power spectrum can be exploited to constrain the astrophysics of reionization, too. However, it is affected by instrumental systematics and observational uncertainties that are comparable to the effect of patchy reionization. For the interest of theoretical understanding, we consider an ``ideal'' survey in which one can effectively separate the astrophysics of reionization completely from all other systematic errors that are extracted before obtaining $P_{F}^{\rm 1D}$. 
Based on the performance of eBOSS, we explore the forecast of reionization parameters using the \lya 1D power spectrum. We find, as speculated in MM20, that splitting the analysis into a higher redshift component and a lower one is a helpful strategy, wherein the high-z observations are useful in constraining the reionization parameters, while the low-z information can focus on cosmology. However, this sweet spot gets compromised slightly because of lower number of available lines of sights with increasing redshifts. In comparison, the 3D forecast does not show this behavior due to a combination of less line of sights increasing the noise level and the choice of treating the aliasing term as pure noise. 



Due to limitations in our hybrid approach and forecasting strategy, our results should be taken as an estimate of the expected performance by DESI to extract the astrophysics of reionization. The accuracy of our forecast can be improved by accounting for better modeling of the IGM, refining the forecasting strategy, and performing a Bayesian inference of the astrophysics of reionization. 
However, as a first forecast in this regard, our results are encouraging \emph{per se}, with only account of the effect on the 3D \lya power spectrum (or only accounting for the effect on the 1D power spectrum). The forecast errors are likely to be further reduced by combining other observational probes of reionization.

\section*{Acknowledgements}
We thank the anonymous referee for insightful comments and suggestions. We thank the DESI collaboration for providing the quasar luminosity function and g-band spectrograph information needed to make a realistic forecast. We are also grateful to Andreu Font-Ribera, Patrick McDonald, Matthew Pieri, Zheng Cai, Bohua Li, Jiachen Jiang, and Shifan Zuo for useful comments and discussions. 
This work is supported by National SKA Program of China (grant No.~2020SKA0110401), NSFC (grant No. 12050410236, 11821303), and National Key R\&D Program of China (grant No.~2018YFA0404502). PMC was supported by the Tsinghua Shui Mu Scholarship.

\section*{Data availability}
The data underlying this article will be shared on reasonable request to the corresponding authors.



\typeout{}

\bibliographystyle{mnras}
\bibliography{fur_DESI} 

\begin{thebibliography}{}
\makeatletter
\relax
\def\mn@urlcharsother{\let\do\@makeother \do\$\do\&\do\#\do\^\do\_\do\%\do\~}
\def\mn@doi{\begingroup\mn@urlcharsother \@ifnextchar [ {\mn@doi@}
  {\mn@doi@[]}}
\def\mn@doi@[#1]#2{\def\@tempa{#1}\ifx\@tempa\@empty \href
  {http://dx.doi.org/#2} {doi:#2}\else \href {http://dx.doi.org/#2} {#1}\fi
  \endgroup}
\def\mn@eprint#1#2{\mn@eprint@#1:#2::\@nil}
\def\mn@eprint@arXiv#1{\href {http://arxiv.org/abs/#1} {{\tt arXiv:#1}}}
\def\mn@eprint@dblp#1{\href {http://dblp.uni-trier.de/rec/bibtex/#1.xml}
  {dblp:#1}}
\def\mn@eprint@#1:#2:#3:#4\@nil{\def\@tempa {#1}\def\@tempb {#2}\def\@tempc
  {#3}\ifx \@tempc \@empty \let \@tempc \@tempb \let \@tempb \@tempa \fi \ifx
  \@tempb \@empty \def\@tempb {arXiv}\fi \@ifundefined
  {mn@eprint@\@tempb}{\@tempb:\@tempc}{\expandafter \expandafter \csname
  mn@eprint@\@tempb\endcsname \expandafter{\@tempc}}}

\bibitem[\protect\citeauthoryear{{Arinyo-i-Prats}, {Miralda-Escud{\'e}}, {Viel}
   \& {Cen}}{{Arinyo-i-Prats} et~al.}{2015}]{2015JCAP...12..017A}
{Arinyo-i-Prats} A.,  {Miralda-Escud{\'e}} J.,  {Viel} M.,   {Cen} R.,  2015,
  \mn@doi [\jcap] {10.1088/1475-7516/2015/12/017}, \href
  {https://ui.adsabs.harvard.edu/abs/2015JCAP...12..017A} {2015, 017}

\bibitem[\protect\citeauthoryear{{Ba{\~n}ados} et~al.,}{{Ba{\~n}ados}
  et~al.}{2018}]{2018Natur.553..473B}
{Ba{\~n}ados} E.,  et~al., 2018, \mn@doi [\nat] {10.1038/nature25180}, \href
  {https://ui.adsabs.harvard.edu/abs/2018Natur.553..473B} {553, 473}

\bibitem[\protect\citeauthoryear{{Blas}, {Lesgourgues}  \& {Tram}}{{Blas}
  et~al.}{2011}]{2011JCAP...07..034B}
{Blas} D.,  {Lesgourgues} J.,   {Tram} T.,  2011, \mn@doi [\jcap]
  {10.1088/1475-7516/2011/07/034}, \href
  {https://ui.adsabs.harvard.edu/abs/2011JCAP...07..034B} {2011, 034}

\bibitem[\protect\citeauthoryear{{Cai} et~al.,}{{Cai}
  et~al.}{2017}]{2017ApJ...839..131C}
{Cai} Z.,  et~al., 2017, \mn@doi [\apj] {10.3847/1538-4357/aa6a1a}, \href
  {https://ui.adsabs.harvard.edu/abs/2017ApJ...839..131C} {839, 131}

\bibitem[\protect\citeauthoryear{{Chabanier} et~al.,}{{Chabanier}
  et~al.}{2019}]{2019JCAP...07..017C}
{Chabanier} S.,  et~al., 2019, \mn@doi [\jcap] {10.1088/1475-7516/2019/07/017},
  \href {https://ui.adsabs.harvard.edu/abs/2019JCAP...07..017C} {2019, 017}

\bibitem[\protect\citeauthoryear{{Chabanier}, {Bournaud}, {Dubois},
  {Palanque-Delabrouille}, {Y{\`e}che}, {Armengaud}, {Peirani}  \&
  {Beckmann}}{{Chabanier} et~al.}{2020}]{2020MNRAS.495.1825C}
{Chabanier} S.,  {Bournaud} F.,  {Dubois} Y.,  {Palanque-Delabrouille} N.,
  {Y{\`e}che} C.,  {Armengaud} E.,  {Peirani} S.,   {Beckmann} R.,  2020,
  \mn@doi [\mnras] {10.1093/mnras/staa1242}, \href
  {https://ui.adsabs.harvard.edu/abs/2020MNRAS.495.1825C} {495, 1825}

\bibitem[\protect\citeauthoryear{{Croft}, {Romeo}  \& {Metcalf}}{{Croft}
  et~al.}{2018a}]{2018MNRAS.477.1814C}
{Croft} R. A.~C.,  {Romeo} A.,   {Metcalf} R.~B.,  2018a, \mn@doi [\mnras]
  {10.1093/mnras/sty650}, \href
  {https://ui.adsabs.harvard.edu/abs/2018MNRAS.477.1814C} {477, 1814}

\bibitem[\protect\citeauthoryear{{Croft}, {Miralda-Escud{\'e}}, {Zheng},
  {Blomqvist}  \& {Pieri}}{{Croft} et~al.}{2018b}]{2018MNRAS.481.1320C}
{Croft} R. A.~C.,  {Miralda-Escud{\'e}} J.,  {Zheng} Z.,  {Blomqvist} M.,
  {Pieri} M.,  2018b, \mn@doi [\mnras] {10.1093/mnras/sty2302}, \href
  {https://ui.adsabs.harvard.edu/abs/2018MNRAS.481.1320C} {481, 1320}

\bibitem[\protect\citeauthoryear{{DESI Collaboration} et~al.,}{{DESI
  Collaboration} et~al.}{2016}]{2016arXiv161100036D}
{DESI Collaboration} et~al., 2016, preprint, \href
  {http://adsabs.harvard.edu/abs/2016arXiv161100036D} {} (\mn@eprint {arXiv}
  {1611.00036})

\bibitem[\protect\citeauthoryear{{Davies} et~al.,}{{Davies}
  et~al.}{2018}]{2018ApJ...864..142D}
{Davies} F.~B.,  et~al., 2018, \mn@doi [\apj] {10.3847/1538-4357/aad6dc}, \href
  {https://ui.adsabs.harvard.edu/abs/2018ApJ...864..142D} {864, 142}

\bibitem[\protect\citeauthoryear{{Font-Ribera} et~al.,}{{Font-Ribera}
  et~al.}{2013}]{2013JCAP...05..018F}
{Font-Ribera} A.,  et~al., 2013, \mn@doi [\jcap]
  {10.1088/1475-7516/2013/05/018}, \href
  {https://ui.adsabs.harvard.edu/abs/2013JCAP...05..018F} {2013, 018}

\bibitem[\protect\citeauthoryear{{Font-Ribera}, {McDonald}, {Mostek}, {Reid},
  {Seo}  \& {Slosar}}{{Font-Ribera} et~al.}{2014}]{2014JCAP...05..023F}
{Font-Ribera} A.,  {McDonald} P.,  {Mostek} N.,  {Reid} B.~A.,  {Seo} H.-J.,
  {Slosar} A.,  2014, \mn@doi [\jcap] {10.1088/1475-7516/2014/05/023}, \href
  {https://ui.adsabs.harvard.edu/abs/2014JCAP...05..023F} {2014, 023}

\bibitem[\protect\citeauthoryear{{Greig} \& {Mesinger}}{{Greig} \&
  {Mesinger}}{2017a}]{2017MNRAS.465.4838G}
{Greig} B.,  {Mesinger} A.,  2017a, \mn@doi [\mnras] {10.1093/mnras/stw3026},
  \href {https://ui.adsabs.harvard.edu/abs/2017MNRAS.465.4838G} {465, 4838}

\bibitem[\protect\citeauthoryear{{Greig} \& {Mesinger}}{{Greig} \&
  {Mesinger}}{2017b}]{2017MNRAS.472.2651G}
{Greig} B.,  {Mesinger} A.,  2017b, \mn@doi [\mnras] {10.1093/mnras/stx2118},
  \href {https://ui.adsabs.harvard.edu/abs/2017MNRAS.472.2651G} {472, 2651}

\bibitem[\protect\citeauthoryear{{Greig}, {Mesinger}, {Haiman}  \&
  {Simcoe}}{{Greig} et~al.}{2017}]{2017MNRAS.466.4239G}
{Greig} B.,  {Mesinger} A.,  {Haiman} Z.,   {Simcoe} R.~A.,  2017, \mn@doi
  [\mnras] {10.1093/mnras/stw3351}, \href
  {https://ui.adsabs.harvard.edu/abs/2017MNRAS.466.4239G} {466, 4239}

\bibitem[\protect\citeauthoryear{{Greig}, {Mesinger}  \& {Ba{\~n}ados}}{{Greig}
  et~al.}{2019}]{2019MNRAS.484.5094G}
{Greig} B.,  {Mesinger} A.,   {Ba{\~n}ados} E.,  2019, \mn@doi [\mnras]
  {10.1093/mnras/stz230}, \href
  {https://ui.adsabs.harvard.edu/abs/2019MNRAS.484.5094G} {484, 5094}

\bibitem[\protect\citeauthoryear{{Hirata}}{{Hirata}}{2018}]{2018MNRAS.474.2173H}
{Hirata} C.~M.,  2018, \mn@doi [\mnras] {10.1093/mnras/stx2854}, \href
  {https://ui.adsabs.harvard.edu/abs/2018MNRAS.474.2173H} {474, 2173}

\bibitem[\protect\citeauthoryear{{Hoag} et~al.,}{{Hoag}
  et~al.}{2019}]{2019ApJ...878...12H}
{Hoag} A.,  et~al., 2019, \mn@doi [\apj] {10.3847/1538-4357/ab1de7}, \href
  {https://ui.adsabs.harvard.edu/abs/2019ApJ...878...12H} {878, 12}

\bibitem[\protect\citeauthoryear{{Keating}, {Weinberger}, {Kulkarni},
  {Haehnelt}, {Chardin}  \& {Aubert}}{{Keating}
  et~al.}{2020}]{2020MNRAS.491.1736K}
{Keating} L.~C.,  {Weinberger} L.~H.,  {Kulkarni} G.,  {Haehnelt} M.~G.,
  {Chardin} J.,   {Aubert} D.,  2020, \mn@doi [\mnras] {10.1093/mnras/stz3083},
  \href {https://ui.adsabs.harvard.edu/abs/2020MNRAS.491.1736K} {491, 1736}

\bibitem[\protect\citeauthoryear{{Kulkarni}, {Keating}, {Haehnelt}, {Bosman},
  {Puchwein}, {Chardin}  \& {Aubert}}{{Kulkarni}
  et~al.}{2019}]{2019MNRAS.485L..24K}
{Kulkarni} G.,  {Keating} L.~C.,  {Haehnelt} M.~G.,  {Bosman} S. E.~I.,
  {Puchwein} E.,  {Chardin} J.,   {Aubert} D.,  2019, \mn@doi [\mnras]
  {10.1093/mnrasl/slz025}, \href
  {https://ui.adsabs.harvard.edu/abs/2019MNRAS.485L..24K} {485, L24}

\bibitem[\protect\citeauthoryear{{Lidz}, {Chang}, {Mas-Ribas}  \& {Sun}}{{Lidz}
  et~al.}{2021}]{2021arXiv210502293L}
{Lidz} A.,  {Chang} T.-C.,  {Mas-Ribas} L.,   {Sun} G.,  2021, arXiv e-prints,
  \href {https://ui.adsabs.harvard.edu/abs/2021arXiv210502293L} {p.
  arXiv:2105.02293}

\bibitem[\protect\citeauthoryear{{Mason}, {Treu}, {Dijkstra}, {Mesinger},
  {Trenti}, {Pentericci}, {de Barros}  \& {Vanzella}}{{Mason}
  et~al.}{2018}]{2018ApJ...856....2M}
{Mason} C.~A.,  {Treu} T.,  {Dijkstra} M.,  {Mesinger} A.,  {Trenti} M.,
  {Pentericci} L.,  {de Barros} S.,   {Vanzella} E.,  2018, \mn@doi [\apj]
  {10.3847/1538-4357/aab0a7}, \href
  {https://ui.adsabs.harvard.edu/abs/2018ApJ...856....2M} {856, 2}

\bibitem[\protect\citeauthoryear{{Mason} et~al.,}{{Mason}
  et~al.}{2019}]{2019MNRAS.485.3947M}
{Mason} C.~A.,  et~al., 2019, \mn@doi [\mnras] {10.1093/mnras/stz632}, \href
  {https://ui.adsabs.harvard.edu/abs/2019MNRAS.485.3947M} {485, 3947}

\bibitem[\protect\citeauthoryear{{McDonald}}{{McDonald}}{2003}]{2003ApJ...585...34M}
{McDonald} P.,  2003, \mn@doi [\apj] {10.1086/345945}, \href
  {https://ui.adsabs.harvard.edu/abs/2003ApJ...585...34M} {585, 34}

\bibitem[\protect\citeauthoryear{{McDonald} \& {Eisenstein}}{{McDonald} \&
  {Eisenstein}}{2007}]{2007PhRvD..76f3009M}
{McDonald} P.,  {Eisenstein} D.~J.,  2007, \mn@doi [\prd]
  {10.1103/PhysRevD.76.063009}, \href
  {https://ui.adsabs.harvard.edu/abs/2007PhRvD..76f3009M} {76, 063009}

\bibitem[\protect\citeauthoryear{{McDonald} et~al.,}{{McDonald}
  et~al.}{2005}]{2005ApJ...635..761M}
{McDonald} P.,  et~al., 2005, \mn@doi [\apj] {10.1086/497563}, \href
  {https://ui.adsabs.harvard.edu/abs/2005ApJ...635..761M} {635, 761}

\bibitem[\protect\citeauthoryear{{McGreer}, {Mesinger}  \&
  {D'Odorico}}{{McGreer} et~al.}{2015}]{2015MNRAS.447..499M}
{McGreer} I.~D.,  {Mesinger} A.,   {D'Odorico} V.,  2015, \mn@doi [\mnras]
  {10.1093/mnras/stu2449}, \href
  {https://ui.adsabs.harvard.edu/abs/2015MNRAS.447..499M} {447, 499}

\bibitem[\protect\citeauthoryear{{McQuinn}}{{McQuinn}}{2016}]{2016ARA&A..54..313M}
{McQuinn} M.,  2016, \mn@doi [\araa] {10.1146/annurev-astro-082214-122355},
  \href {http://adsabs.harvard.edu/abs/2016ARA%26A..54..313M} {54, 313}

\bibitem[\protect\citeauthoryear{{Mesinger}, {Furlanetto}  \& {Cen}}{{Mesinger}
  et~al.}{2011}]{2011MNRAS.411..955M}
{Mesinger} A.,  {Furlanetto} S.,   {Cen} R.,  2011, \mn@doi [\mnras]
  {10.1111/j.1365-2966.2010.17731.x}, \href
  {https://ui.adsabs.harvard.edu/abs/2011MNRAS.411..955M} {411, 955}

\bibitem[\protect\citeauthoryear{{Mesinger}, {Aykutalp}, {Vanzella},
  {Pentericci}, {Ferrara}  \& {Dijkstra}}{{Mesinger}
  et~al.}{2015}]{2015MNRAS.446..566M}
{Mesinger} A.,  {Aykutalp} A.,  {Vanzella} E.,  {Pentericci} L.,  {Ferrara} A.,
    {Dijkstra} M.,  2015, \mn@doi [\mnras] {10.1093/mnras/stu2089}, \href
  {https://ui.adsabs.harvard.edu/abs/2015MNRAS.446..566M} {446, 566}

\bibitem[\protect\citeauthoryear{{Mirocha}, {La Plante}  \& {Liu}}{{Mirocha}
  et~al.}{2020}]{2020arXiv201209189M}
{Mirocha} J.,  {La Plante} P.,   {Liu} A.,  2020, arXiv e-prints, \href
  {https://ui.adsabs.harvard.edu/abs/2020arXiv201209189M} {p. arXiv:2012.09189}

\bibitem[\protect\citeauthoryear{{Mirocha}, {Lamarre}  \& {Liu}}{{Mirocha}
  et~al.}{2021}]{2021MNRAS.504.1555M}
{Mirocha} J.,  {Lamarre} H.,   {Liu} A.,  2021, \mn@doi [\mnras]
  {10.1093/mnras/stab949}, \href
  {https://ui.adsabs.harvard.edu/abs/2021MNRAS.504.1555M} {504, 1555}

\bibitem[\protect\citeauthoryear{{Montero-Camacho} \& {Mao}}{{Montero-Camacho}
  \& {Mao}}{2020}]{2020MNRAS.499.1640M}
{Montero-Camacho} P.,  {Mao} Y.,  2020, \mn@doi [\mnras]
  {10.1093/mnras/staa2918}, \href
  {https://ui.adsabs.harvard.edu/abs/2020MNRAS.499.1640M} {499, 1640}

\bibitem[\protect\citeauthoryear{{Montero-Camacho}, {Hirata}, {Martini}  \&
  {Honscheid}}{{Montero-Camacho} et~al.}{2019}]{2019MNRAS.487.1047M}
{Montero-Camacho} P.,  {Hirata} C.~M.,  {Martini} P.,   {Honscheid} K.,  2019,
  \mn@doi [\mnras] {10.1093/mnras/stz1388}, \href
  {https://ui.adsabs.harvard.edu/abs/2019MNRAS.487.1047M} {487, 1047}

\bibitem[\protect\citeauthoryear{{Nasir} \& {D'Aloisio}}{{Nasir} \&
  {D'Aloisio}}{2020}]{2020MNRAS.494.3080N}
{Nasir} F.,  {D'Aloisio} A.,  2020, \mn@doi [\mnras] {10.1093/mnras/staa894},
  \href {https://ui.adsabs.harvard.edu/abs/2020MNRAS.494.3080N} {494, 3080}

\bibitem[\protect\citeauthoryear{{O{\~n}orbe}, {Davies}, {Luki{\'c}}, {},
  {Hennawi}  \& {Sorini}}{{O{\~n}orbe} et~al.}{2019}]{2019MNRAS.486.4075O}
{O{\~n}orbe} J.,  {Davies} F.~B.,  {Luki{\'c}} {} Z.,  {Hennawi} J.~F.,
  {Sorini} D.,  2019, \mn@doi [\mnras] {10.1093/mnras/stz984}, \href
  {https://ui.adsabs.harvard.edu/abs/2019MNRAS.486.4075O} {486, 4075}

\bibitem[\protect\citeauthoryear{{Ouchi} et~al.,}{{Ouchi}
  et~al.}{2010}]{2010ApJ...723..869O}
{Ouchi} M.,  et~al., 2010, \mn@doi [\apj] {10.1088/0004-637X/723/1/869}, \href
  {https://ui.adsabs.harvard.edu/abs/2010ApJ...723..869O} {723, 869}

\bibitem[\protect\citeauthoryear{{Palanque-Delabrouille}
  et~al.,}{{Palanque-Delabrouille} et~al.}{2013a}]{2013A&A...551A..29P}
{Palanque-Delabrouille} N.,  et~al., 2013a, \mn@doi [\aap]
  {10.1051/0004-6361/201220379}, \href
  {https://ui.adsabs.harvard.edu/abs/2013A&A...551A..29P} {551, A29}

\bibitem[\protect\citeauthoryear{{Palanque-Delabrouille}
  et~al.,}{{Palanque-Delabrouille} et~al.}{2013b}]{2013A&A...559A..85P}
{Palanque-Delabrouille} N.,  et~al., 2013b, \mn@doi [\aap]
  {10.1051/0004-6361/201322130}, \href
  {https://ui.adsabs.harvard.edu/abs/2013A&A...559A..85P} {559, A85}

\bibitem[\protect\citeauthoryear{{Park}, {Mesinger}, {Greig}  \&
  {Gillet}}{{Park} et~al.}{2019}]{2019MNRAS.484..933P}
{Park} J.,  {Mesinger} A.,  {Greig} B.,   {Gillet} N.,  2019, \mn@doi [\mnras]
  {10.1093/mnras/stz032}, \href
  {https://ui.adsabs.harvard.edu/abs/2019MNRAS.484..933P} {484, 933}

\bibitem[\protect\citeauthoryear{{Park}, {Shapiro}, {Ahn}, {Yoshida}  \&
  {Hirano}}{{Park} et~al.}{2021}]{2021ApJ...908...96P}
{Park} H.,  {Shapiro} P.~R.,  {Ahn} K.,  {Yoshida} N.,   {Hirano} S.,  2021,
  \mn@doi [\apj] {10.3847/1538-4357/abd7f4}, \href
  {https://ui.adsabs.harvard.edu/abs/2021ApJ...908...96P} {908, 96}

\bibitem[\protect\citeauthoryear{{Pieri} et~al.,}{{Pieri}
  et~al.}{2016}]{2016sf2a.conf..259P}
{Pieri} M.~M.,  et~al., 2016, in {Reyl{\'e}} C.,  {Richard} J.,  {Cambr{\'e}sy}
  L.,  {Deleuil} M.,  {P{\'e}contal} E.,  {Tresse} L.,   {Vauglin} I.,  eds,
  SF2A-2016: Proceedings of the Annual meeting of the French Society of
  Astronomy and Astrophysics. pp 259--266 (\mn@eprint {arXiv} {1611.09388})

\bibitem[\protect\citeauthoryear{{Planck Collaboration} et~al.,}{{Planck
  Collaboration} et~al.}{2020}]{2020A&A...641A...6P}
{Planck Collaboration} et~al., 2020, \mn@doi [\aap]
  {10.1051/0004-6361/201833910}, \href
  {https://ui.adsabs.harvard.edu/abs/2020A&A...641A...6P} {641, A6}

\bibitem[\protect\citeauthoryear{{Pontzen}}{{Pontzen}}{2014}]{2014PhRvD..89h3010P}
{Pontzen} A.,  2014, \mn@doi [\prd] {10.1103/PhysRevD.89.083010}, \href
  {https://ui.adsabs.harvard.edu/abs/2014PhRvD..89h3010P} {89, 083010}

\bibitem[\protect\citeauthoryear{{Richard} et~al.,}{{Richard}
  et~al.}{2019}]{2019Msngr.175...50R}
{Richard} J.,  et~al., 2019, \mn@doi [The Messenger] {10.18727/0722-6691/5127},
  \href {https://ui.adsabs.harvard.edu/abs/2019Msngr.175...50R} {175, 50}

\bibitem[\protect\citeauthoryear{{Sobacchi} \& {Mesinger}}{{Sobacchi} \&
  {Mesinger}}{2015}]{2015MNRAS.453.1843S}
{Sobacchi} E.,  {Mesinger} A.,  2015, \mn@doi [\mnras] {10.1093/mnras/stv1751},
  \href {https://ui.adsabs.harvard.edu/abs/2015MNRAS.453.1843S} {453, 1843}

\bibitem[\protect\citeauthoryear{{Springel}}{{Springel}}{2005}]{2005MNRAS.364.1105S}
{Springel} V.,  2005, \mn@doi [\mnras] {10.1111/j.1365-2966.2005.09655.x},
  \href {http://adsabs.harvard.edu/abs/2005MNRAS.364.1105S} {364, 1105}

\bibitem[\protect\citeauthoryear{{Upton Sanderbeck} \& {Bird}}{{Upton
  Sanderbeck} \& {Bird}}{2020}]{2020MNRAS.496.4372U}
{Upton Sanderbeck} P.,  {Bird} S.,  2020, \mn@doi [\mnras]
  {10.1093/mnras/staa1850}, \href
  {https://ui.adsabs.harvard.edu/abs/2020MNRAS.496.4372U} {496, 4372}

\bibitem[\protect\citeauthoryear{{Wolz}, {Kilbinger}, {Weller}  \&
  {Giannantonio}}{{Wolz} et~al.}{2012}]{2012JCAP...09..009W}
{Wolz} L.,  {Kilbinger} M.,  {Weller} J.,   {Giannantonio} T.,  2012, \mn@doi
  [\jcap] {10.1088/1475-7516/2012/09/009}, \href
  {https://ui.adsabs.harvard.edu/abs/2012JCAP...09..009W} {2012, 009}

\bibitem[\protect\citeauthoryear{{Wu}, {McQuinn}, {Kannan}, {D'Aloisio},
  {Bird}, {Marinacci}, {Dav{\'e}}  \& {Hernquist}}{{Wu}
  et~al.}{2019}]{2019MNRAS.490.3177W}
{Wu} X.,  {McQuinn} M.,  {Kannan} R.,  {D'Aloisio} A.,  {Bird} S.,  {Marinacci}
  F.,  {Dav{\'e}} R.,   {Hernquist} L.,  2019, \mn@doi [\mnras]
  {10.1093/mnras/stz2807}, \href
  {https://ui.adsabs.harvard.edu/abs/2019MNRAS.490.3177W} {490, 3177}

\bibitem[\protect\citeauthoryear{{Y{\`e}che} et~al.,}{{Y{\`e}che}
  et~al.}{2020}]{2020RNAAS...4..179Y}
{Y{\`e}che} C.,  et~al., 2020, \mn@doi [Research Notes of the American
  Astronomical Society] {10.3847/2515-5172/abc01a}, \href
  {https://ui.adsabs.harvard.edu/abs/2020RNAAS...4..179Y} {4, 179}

\makeatother
\end{thebibliography}







\bsp	
\label{lastpage}
\end{document}